\documentclass[leqno]{amsart}

\usepackage[foot]{amsaddr}

\usepackage{color}
\usepackage{geometry}

\usepackage[numbers]{natbib}
\usepackage{amsmath}
\usepackage{mathtools}

\usepackage{subfiles}
\usepackage{graphicx}
\usepackage[section]{placeins}
\usepackage{hyperref}
\usepackage{stmaryrd}
\usepackage{amsmath,amssymb,amsfonts,amsthm,textcomp}
\usepackage{mathtools}
\usepackage{tensor}
\usepackage{epstopdf}
\usepackage{multicol}
\usepackage{empheq}
\usepackage{tikz,pgfplots}
\usetikzlibrary{calc}
\usepackage{enumitem}
 \newcommand{\lJump}{[\![}
\newcommand{\rJump}{]\!]}

\newtheorem{remark}{Remark}
\newtheorem{proposition}{Proposition}
\usepackage[linesnumbered,ruled,vlined]{algorithm2e}
\usepackage{subfigure}
\usepackage{fancyhdr}

\newcommand*\circled[1]{\tikz[baseline=(char.base)]{
		\node[shape=circle,draw,inner sep=2pt] (char) {#1};}}

\graphicspath{{./Figures/}
	{./Figures/Formulation/}
	{./Figures/Intro/}
	{./Figures/NumericalModel/}
	{./Figures/Analytical/} }

\newfont{\tenbfsl}{cmbxti9 scaled 1200}
\newfont{\tenbbb}{msbm10}
\newfont{\svnbbb}{msbm8}

\theoremstyle{remark}

\theoremstyle{definition}

\newcounter{syn}[section] 
\renewcommand{\thesyn}{\arabic{section}.\arabic{syn}}

\makeatletter
\def\threevdots{\mskip+4mu\vbox{\baselineskip2.25\p@ \lineskiplimit\z@
  \kern4.9\p@\hbox{.}\hbox{.}\hbox{.}}\mskip+3.8mu}
\makeatother

\usepackage{hyperref}
\hypersetup{
	colorlinks   = true, %Colours links instead of ugly boxes
	urlcolor     = blue, %Colour for external hyperlinks
	linkcolor    = blue, %Colour of internal links
	citecolor   = red %Colour of citations
}
\begin{document}
\sloppypar
\title{
Compaction band localization in geomaterials: a mechanically consistent failure criterion
}
\author{Roberto J. Cier$^{\flat,\sharp,\ddagger}$}
\email{roberto.cier@pucp.edu.pe (R.J. Cier)}
\address{$^{\flat}$ School of Civil and Mechanical Engineering, Curtin University, Bentley, WA 6102, Australia}
\author{Nicolas A. Labanda$^{\natural,\ddagger}$}
\email{nlabanda@facet.unt.edu.ar (N.A. Labanda)}
\address{$^{\natural}$ School of Electrical Engineering, Computing and Mathematical Sciences, Curtin University, Bentley, WA 6102, Australia}
\address{$^{\sharp}$ CSIRO Mineral Resources, Kensington, WA 6151, Australia}
\address{$^{\dagger}$ Curtin Institute for Computation, Curtin University, Bentley, WA 6102, Australia, Australia}
\address{$^{\ddagger}$ SRK Consulting (Australasia) Pty Ltd, West Perth, WA 6005, Australia}
%\email{nlabanda@facet.unt.edu.ar (N.A. Labanda)}
\author{Victor M. Calo$^{\natural,\sharp,\dagger}$}
\email{victor.calo@curtin.edu.au (V.M. Calo)}

\date{\today}

 \begin{abstract}
 \noindent
 Compaction bands play a key role in the deformation processes of porous rocks and explain different aspects of physical processes in geological formations. The state-of-the-art description of the localized strains that lead to compaction banding has limitations from the mechanical point of view. Thus, we describe the phenomenon using a consistent axiomatic formulation. We build a viscoplastic model using minimal assumptions; we base our model on six principles to study compaction band localization triggered by viscous effects.  We analyze different stress states to determine the conditions that trigger compaction bands. Laboratory experiments show that a material undergoes different localizations depending on the confinement pressure; thus, we perform a series of numerical experiments that reproduce these phenomena under varying triaxial compression conditions. These simulations use a simple viscoplastic constitutive model for creep based on Perzyna's viscoplasticity and show how confinement changes the localization type for different triaxial tests. Our analysis allows us to describe this transition, band periodicity and spacing, and their dependence on the material parameters.


 \end{abstract}

\maketitle

%\pagestyle{fancy}
%\fancyhf{}
%\rhead{\thepage}
%\lhead{Automatic adaptive time-stepping scheme for hyperbolic partial differential equations}

\tableofcontents                        % Print table of contents

%-------------------------------------------------------------------------------%

 % !TeX spellcheck = en_US
\section{Introduction}
\label{section:Introduction}

Compaction bands are narrow planar zones that appear predominantly in porous rocks with their normals parallel to the maximum principal stress. These bands are highly dense, with low permeability, and have different mineralogical compositions than the rest of the matrix; they appear mostly under compressive states of stress. The compaction bands play a predominant role in geomechanics and earth sciences as many geological features such as faults, folds, boudinage, landslides, and mineralization, to name a few, fall within this type of phenomenon. From an industrial point of view, compaction bands are crucial in the study of unconventional resources, which includes those subjected to volatile conditions, deeper and hotter than ever reached before and in a challenging environment for their extraction~\cite{Regenauer-Lieb2016}. Additionally, compaction bands play an essential role in the energy sector, for example, in shale gas and oil reservoir simulation~\cite{Alevizos2017} and the mineral exploration sector, as those bands can become pathways for mineralising fluids~\cite{Kelka2017, Poulet2017}.

From a theoretical point, compaction bands fall within the family of localization phenomena, along with shear and dilation bands. In solid mechanics, the study of strain localization goes back to Hill~\cite{ Hill1962}, who associated material instabilities with the stationary limit of the accelerating wave velocity in solids. Based on this early work, a series of bifurcation criteria were developed for geomaterials, being the first and most widely used one by Rudnicki and Rice~\cite{ Rudnicki1975}. This framework established a series of conditions for the shear banding inception as a bifurcation problem in brittle rocks under principal compressive stresses. Later, Olsson~\cite{ Olsson1999} extended the bifurcation criterion to compaction banding scenarios, and finally, Issen and Rudnicki~\cite{ Issen2001} formalized the theory for the onset of compaction bands in porous rocks by incorporating a comprehensive bifurcation approach that uses a cap-type yield function. These approaches use under strain-controlled conditions and rely on the concept of the acoustic tensor (dependent on the consistent-tangent constitutive tensor). Alternatively, a simple approach for stress-controlled scenarios was proposed by Vermeer~\cite{ vermeer1982}, limited only to two-dimensional conditions but more straightforward to implement compared to the primal work of Rudnicki and Rice~\cite{ Gutierrez2017}. Nova~\cite{ nova1994} proposed bifurcation conditions for viscoplastic processes derived from the controllability criterion for mixed-mode loading cases. Pisan\`o and di~Prisco~\cite{ pisano2016} introduced a stability criterion for elastoviscoplastic constitutive laws from the spectral analysis of the resulting matrix of an ordinary partial differential equation, derived from expressing a second-order form of the Perzyna's~\cite{ PERZYNA1966243} constitutive equations.

The construction of a constitutive tangent tensor limits the localization evaluation in rate-dependent materials, as classical viscoplastic theories do not include consistency conditions, precluding the recovery of a viscoplastic constitutive tensor. In plasticity theory, the plastic multiplier is a consequence of the constitutive assumptions~\citep{ SIMO1985101} rather than an explicit computation given by ad-hoc definitions. Additionally, studies showed that inconsistent (visco-)plastic multipliers, such as Perzyna's~\citep{ PERZYNA1966243} or Duvaut-Lions'~\citep{ dl1982}, produce loading-unloading stress paths that induce energy dissipation~\citep{ HEERES20021}. Consistent visco-plastic formulations require fundamental definitions to redefine the yield function that becomes time-dependent. Nonetheless, many authors still neglect consistency under viscoplastic scenarios and compute the viscoplastic component inconsistently. A few researchers formulated consistent viscoplastic approaches that allowed them to develop strain localization analyses~\cite{ CAROSIO20007349, Wang1997}, although they only focused on shear banding and used perfect ${J_2~\text{elastoplasticty}}$. 

We seek to overcome some of the shortcomings of the bifurcation analysis of compaction banding for rate-dependent materials. We study, analytically and numerically, the compaction band localization for a class of consistent viscoplastic critical-state-based constitutive models. We theoretically analyze the compaction bands' onset based on a consistent viscoplastic constitutive framework for the Modified Cam-Clay model (MCC)~\cite{ roscoe1968, wood1990}. We focus on well-known loading cases in geomechanics practice. Additionally, we use numerical experiments to corroborate the localization behavior under triaxial compression and show the transition of a sample response from shear banding to multiple compaction banding scenarios by adjusting the confinement pressure in the test system.

We organize the paper as follows: Section~\ref{section:Theory} introduces our theoretical framework, detailing our axiomatic construction and its different components. Section~\ref{section:constitutivetensor} analyzes the compaction banding localization, dealing first with the reconstruction of the visco-plastic constitutive tensor and then analyzing the bifurcation onset from the acoustic tensor in different stress scenarios. Section~\ref{section:numexp} develops a series of numerical experiments carried out using a class of viscoplastic models in line with our theoretical proposal and shows the progression of the compaction bands under triaxial compression states. Finally, we draw conclusions in Section~\ref{section:Conclusions}.
 \section{Theoretical framework}
\label{section:Theory}

\begin{figure}
  \centering
  \includegraphics[width=0.65\textwidth]{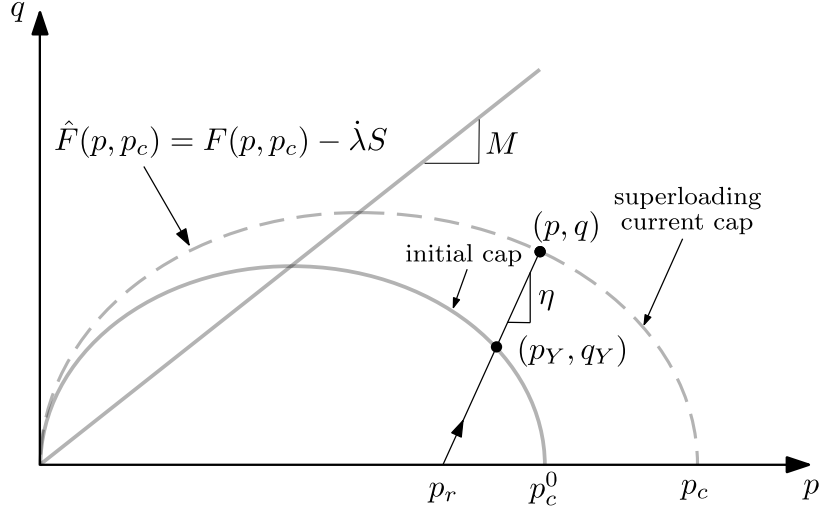} 
  \caption{Problem statement for a modified Cam-Clay-type cap surface.}
  \label{fig:mcc_plot}
\end{figure}

In this paper, we use Cambridge's notation for stress invariants~\cite{ roscoe1958}, that is, for principal effective stresses, the mean and the deviatoric stresses are:
\begin{align}
  p'=\frac{\sigma'_{11}+\sigma'_{22}+\sigma'_{33}}{3},&& q=\frac{1}{\sqrt{2}}{\sqrt{(\sigma'_{11} - \sigma'_{22})^2 + (\sigma'_{22} - \sigma'_{33})^2 + (\sigma'_{22} - \sigma'_{33})^2 + {\sigma'}^{2}_{12} + {\sigma'}^{2}_{23} + {\sigma'}^{2}_{31}}}.
\end{align}
Moreover, we assume the samples undergo straight stress paths; we consider the pair $(p,q)$ follows a known obliquity (stress ratio) $\eta$, such that $q=\eta(p-p_r)$,  where $p_r$ is the reference mean stress. For instance, an isotropically consolidated drained compression (CIDC) triaxial test in geomaterials considers $p_r = \sigma'_3$ and $\eta=3$, whereas in an isotropic compression test $p_r=0$ and $\eta=0$ (see Figure~\ref{fig:mcc_plot} for a sketch of these ideas). This assumption allows us to state the problem exclusively in terms of the mean stresses and the obliquity. 

\subsection{Model statement}

In this work, we propose a framework that can consistently formulate any viscoelastoplastic model by specifying the following \textit{six} features:
\begin{enumerate}[label=(\roman*)]
\item An elastic constitutive behavior: 
  \begin{equation*}
    \sigma'_{ij} = \frac{\partial \psi^{e} \left(  \varepsilon^{e} \right)}{\partial  \varepsilon^{e}_{kl}} = \mathbb{C}^{e}_{ijkl}  \varepsilon^{e}_{kl} \,,
  \end{equation*}
  where $\psi^{e} \left(  \varepsilon^{e} \right)$ is the scalar elastic potential, and $\mathbb{C}^{e}_{ijkl} = \frac{ \partial \psi^{e} \left(  \varepsilon^{e} \right)}{\partial \varepsilon^{e}_{ij} \otimes \varepsilon^{e}_{kl}}$ is the elastic tangent tensor.
\item A kinematic compatibility condition between reversible and irreversible strains: 
  \begin{equation*}
    \dot{\varepsilon}_{ij} = \dot{\varepsilon}^{e}_{ij} + \dot{\varepsilon}^{vp}_{ij}\,. 
  \end{equation*}
\item The existence of an elastic region $E$, bounded by a yield surface $F$.
\item A visco-plastic strain evolution, usually expressed as: 
  \begin{equation*}
    \dot{\varepsilon}^{vp}_{ij} = \dot{\lambda} \frac{\partial G}{\partial \sigma'_{ij}} \, , 
  \end{equation*}
  where $G$ is a plastic potential function.
\item An evolution law for internal variables (hardening/softening rules). In our theory, the only internal variable  is the preconsolidation pressure $p_c$;
\item An overstress or superload surface \textit{active above the yield surface} $F$, that results in a time-dependent yield function $\hat{F}$, with the following a general structure:
  \begin{equation*}
    \hat{F} = F -  \dot{\lambda}  S \,,
  \end{equation*}
  where $S$ depends on $F$ at the current state; this feature distinguishes elastoviscoplastic models from elastoplastic ones, which require only the first five features. Usually, we define $S$ using classical viscoplastic definitions; thus, we follow Perzyna's definition that uses a time-dependent yield function $\hat{F}$ to compute consistency conditions to simulate the viscous effect without spurious dissipation. 
\end{enumerate}

We now introduce a theoretical framework based on the above assumptions to obtain a simple formulation. We define of remaining functions $F$, $G$, $S$ and the time-evolution of the preconsolidation pressure $\dot{p}_{c}$.

\subsection{Elastic behavior}

We define an elastic potential as:
\begin{equation} \label{eq:elasticpot}
  \psi^{e} \left(  \varepsilon^{e} \right) = \frac{1}{2} \left( K_{ur} - \frac{2}{3} G_{ur} \right)  \text{tr}^2 \varepsilon^{e}_{ij} + G_{ur} \varepsilon^{e}_{ij} \varepsilon^{e}_{ij}  \,,
\end{equation}
and consequently, the elastic constitutive tensor reads:
\begin{equation} \label{eq:elasticreg}
  \mathbb{C}^{e}_{ijkl} = \frac{ \partial \psi^{e} \left(  \varepsilon^{e} \right)}{\partial \varepsilon^{e}_{ij} \otimes \varepsilon^{e}_{kl}} =  \left( K_{ur} - \frac{2}{3} G_{ur} \right) {\textbf{I}} \otimes {\textbf{I}} + 2 G_{ur} \mathbb{I} \,,
\end{equation}
where ${\textbf{I}} = \delta_{ij}$ is the second-order tensor identity, and $\mathbb{I} = \frac{1}{2}\left(\delta_{ik} \delta_{jl} + \delta_{il} \delta_{jk}\right)$ represents the fourth-order tensor identity. Two stiffness parameters, the unloading-reloading bulk modulus $K_{ur}$ and the unloading-reloading shear modulus $G_{ur}$  define the elastic response.  These definitions allow for a straightforward computation of the consistent viscoplastic constitutive tensor; then given its definition, we analyze its spectral properties

\subsection{Yield function $F$ and plastic potential function $G$}

Using the modified Cam-Clay model ideas~\citep{ roscoe1968}, we express the yield function $F(p\left( \sigma'_{ij} \right), q, p_c)$ as:
\begin{equation}
  F(p\left( \sigma'_{ij} \right), p_c)=\frac{q^2}{M^2p}+p-p_c=\left(\frac{\eta}{M}\right)^2\frac{(p-p_r)^2}{p}+p-p_c,
\end{equation}
where $M$ represents the slope of the critical state line (CSL), $p_r$ is the reference pressure, and $p_c$ refers to the preconsolidation pressure. Our derivations assume that $(p,q)$ is a post-yield state, which implies that there exists a non-zero visco-plastic deformation. We consider an associative viscoplastic flow; thus, $G \equiv F$.

\subsection{Viscous evolution law $ S $}

Following Perzyna's overstress ideas~\citep{Perzyna1966}, we assume there exists a superloading post-yield surface $\hat{F}(p\left( \sigma'_{ij} \right),q,p_c)\leq0$ at some instant $t$, such that:
\begin{equation}\label{eq:F_hat}
  \hat{F}(p \left( \sigma'_{ij} \right),p_c,\dot{\lambda})=F(p \left( \sigma'_{ij} \right),p_c) - \dot{\lambda}S,
\end{equation}
where $S$ (scaling factor) estimates the overstress with respect to the yield function:
\begin{equation}
  S=\frac{\langle F \left( p \left( \sigma'_{ij} \right) , {p_c} \right) \rangle^{m}}{\mu},
\end{equation}
where $\langle \cdot \rangle$ stands for the Macaulay bracket, $m$ depends on the compression index $\lambda^{*}$, the swelling index $\kappa^{*}$ and the adimensional viscosity parameter $\mu^{*}$, typical material parameters used in critical-state theories for geomaterials. Besides, $\mu = {\mu^{*}/\tau}$ represents the viscosity rate with units ${\text{s}^{-1}}$, measured as the strain produced in a reference time frame $\tau$. For simplicity and without loss of generality, we fix the exponent to $m = 1$.

Our assumptions define the viscoplastic strain rate as:
\begin{equation}\label{eq:eps_vp}
  \dot{\varepsilon}^{vp}_{ij} = \dot{\lambda} {\textbf{N}_{ij}} \, ,
\end{equation}
where ${\textbf{N}_{ij}}$ is the partial derivative of the plastic potential $G$ respect to the effective stress $\sigma'_{ij}$, the (visco)plastic flow. By associativity, we define the plastic potential as the yield function $F$, leading in the following definition of the plastic flux:
\begin{equation} \label{eq:fluxplastic}
  {\textbf{N}_{ij}} := \frac{\partial {F}}{\partial \sigma'_{ij}} \, .
\end{equation}
Similarly, for $\hat{F}$, we express its derivative as  $\hat{\textbf{N}} = \frac{\partial \hat{F}}{\partial \sigma'_{ij}}$. Additionally, we split the flux $\textbf{N}$, into two orthogonal directions, the deviatoric and volumetric components (i.e., $\textbf{N}_\text{d}$ and $N_\text{v}$):
\begin{equation}\label{eq:evpNdNv}
  \dot{\boldsymbol{\varepsilon}}^{vp} = \dot{\lambda} \textbf{N} = \dot{\lambda} (\textbf{N}_\text{d} + N_\text{v} \textbf{I}) \, ;
\end{equation}
the deviatoric and volumetric strain rates then become:
\begin{equation}\label{eq:flowsplit}
  \dot{\boldsymbol{\varepsilon}}^{vp}_\text{d} = \dot{\lambda} \textbf{N}_d \quad ; \quad \dot{\varepsilon}^{vp}_\text{v} = \dot{\lambda} N_\text{v} \, .
\end{equation}
The above definition of the plastic flux follows the classical critical-state assumptions in the Modified Cam-Clay model, and defines a compressive zone (${\dot{\boldsymbol{\varepsilon}}^{vp}_\text{v} < 0}$) for ${p>p_c/2}$, a dilatant zone (${\dot{\boldsymbol{\varepsilon}}^{vp}_\text{v} > 0}$) for ${p<p_c/2}$ (supercritical states) and an isochoric zone (${\dot{\boldsymbol{\varepsilon}}^{vp}_\text{v} = 0}$) at ${p = p_c/2}$ (critical state).

\subsection{Volumetric hardening law: preconsolidation pressure evolution}

A hardening law for the preconsolidation stress increment $\dot{p_c}$ in terms of the viscoplastic strain-rate volumetric contribution $\dot{\varepsilon}_\text{v}^{vp}$ is:
\begin{equation}\label{eq:hlaw}
  \dot{p_c}= H \dot{\varepsilon}_\text{v}^{vp}  = H \dot{\lambda}  N_\text{v} \, ,
\end{equation}
where $H$ represents the hardening parameter that follows classical critical-state-based models:
\begin{equation}\label{eq:Hdef}
  H = \frac{p_c}{\lambda^{*} - \kappa^{*}},
\end{equation}
and 
\begin{equation}\label{eq:Nv}
  N_\text{v} = 1 - \left({\frac{q}{M p}}\right)^{2} = 1 - \left({\frac{\eta \left( p - p_r \right)}{M p}}\right)^{2}  \,,
\end{equation}
is the volumetric contribution of the plastic flow as in~\eqref{eq:flowsplit}.

\subsection{Visco-plastic constraint and explicit preconsolidation evolution}

These conditions are equivalent to the Prager's consistency condition in terms of $\hat{F}(p,p_c,\dot{\lambda})$, which reads:
\begin{equation}\label{eq:consistenc}
  \dot{\hat{F}}(p \left( \sigma'_{ij} \right) ,p_c,\dot{\lambda})=\frac{\partial \hat{F}}{\partial \sigma'_{ij}} \dot{\sigma}'_{ij} + \frac{\partial \hat{F}}{\partial p_c} \dot{p_c} + \frac{\partial \hat{F}}{\partial \dot {\lambda}} \ddot{\lambda}= 0.
\end{equation}
Using the volumetric hardening law~\eqref{eq:hlaw}, we rewrite~\eqref{eq:consistenc} as:
\begin{equation}\label{eq:consistenc_b}
  \dot{\hat{F}}(p \left( \sigma'_{ij} \right),p_c,\dot{\lambda})=\frac{\partial \hat{F}}{\partial \sigma'_{ij}} \dot{\sigma}'_{ij} + \frac{\partial \hat{F}}{\partial p_c} \dot{\lambda} H N_\text{v}  + \frac{\partial \hat{F}}{\partial \dot {\lambda}} \ddot{\lambda}= 0.
\end{equation}
where the partial derivatives are:
\begin{equation}
  \begin{aligned}
    \displaystyle	\frac{\partial \hat{F}}{\partial \sigma'_{ij}} =\displaystyle  {\textbf{N}}_{ij} \,&&
    \displaystyle	\frac{\partial \hat{F}}{\partial p_c} = \displaystyle -1  ,&&
    \displaystyle	\frac{\partial \hat{F}}{\partial \dot {\lambda}} = -S.
  \end{aligned}
\end{equation}

The consistency condition allows us to compute the exact solution of the plastic multiplier and the consistent tangent constitutive tensor.
%Inserting these partial derivatives,~\eqref{eq:consistenc_b} reduces to:
%\begin{equation}\label{eq:consistenc_c}
%	\dot{\hat{F}}(p\left( \sigma'_{ij} \right),p_c)={\textbf{N}}_{ij} \dot{\sigma}'_{ij}  -  H N_\text{v}   \dot{\lambda} -S\ddot{\lambda}= 0 \,.
%\end{equation}
%Thus, we rewrite the consistency condition introducing an auxiliary variable $\chi=\dot{\lambda}$. Then,~\eqref{eq:consistenc_c} is given by:
%\begin{equation}\label{eq:consistency_2}
%	\dot{\hat{F}}(p\left( \sigma'_{ij} \right),p_c)={\textbf{N}}_{ij} \dot{\sigma}'_{ij}  -  H  N_\text{v}  \chi -S\dot{\chi}= 0 \,,
%\end{equation}
%
%The last equation has the structure of first order ordinary equation in terms of $\chi$, with analytical solution of the following way:
%\begin{equation}\label{eq:chi_sol}
%	\chi = C_0 \exp\left(-\frac{H N_\text{v}}{S}t\right) + \frac{1}{H}{\textbf{N}}_{ij} \dot{\sigma}'_{ij} \,.
%\end{equation} 

%A numerical integration will be used to solve the consistency equation, which will be explained in Section \ref{section:implementation}.

 \section{Compaction banding localization analysis}\label{section:constitutivetensor}
\subsection{Viscoplastic constitutive tensor recovery}

We recover the viscoplastic constitutive tensor following similar previous approaches  \cite{Wang1997,CAROSIO20007349}. For this task, we assume a multi-axial stress compatibility. Then, starting from Eq. \eqref{eq:consistenc_b}, the consistency condition reads:
\begin{align}
  0&=\dot{\hat{F}}(\sigma_{ij},p_c,\dot{\lambda}) \nonumber\\
   &=\frac{\partial \hat{F}}{\partial \sigma_{ij}} \dot{\sigma}_{ij} + \frac{\partial \hat{F}}{\partial p_c} \dot{\lambda} H N_\text{v} + \frac{\partial \hat{F}}{\partial \dot {\lambda}} \ddot{\lambda} \nonumber\\
   &=\frac{\partial \hat{F}}{\partial \sigma_{ij}} \mathbb{C}^{e}_{ijkl} \left( \dot{\varepsilon}_{ij}  - \dot{\varepsilon}^{vp}_{ij}  \right) + \frac{\partial \hat{F}}{\partial p_c} \dot{\lambda} H  N_\text{v}  + \frac{\partial \hat{F}}{\partial \dot{\lambda}} \ddot{\lambda}	\nonumber\\
   &=\frac{\partial \hat{F}}{\partial \sigma_{ij}} \mathbb{C}^{e}_{ijkl} \left( \dot{\varepsilon}_{ij}  - \dot{\lambda}  {\textbf{N}_{ij}} \right) + \frac{\partial \hat{F}}{\partial p_c} \dot{\lambda} H N_\text{v}  - S \ddot{\lambda} \nonumber\\
   &= {\textbf{N}} : \mathbb{C}^{e}: \dot{\varepsilon} - \left( {\textbf{N}} : \mathbb{C}^{e}: {\textbf{N}} -  \frac{\partial \hat{F}}{\partial p_c} H N_\text{v}  \right) \dot{\lambda} - S  \ddot{\lambda} \nonumber\\
   & = a + b \dot{\lambda} + c  \ddot{\lambda} \label{eq:consistenc_multiax}\, ,
\end{align}
which results in a first-order differential equation, with exact solution. We parametrize the overstress function $S$ in terms of the trial pressure $p_{0}$ and the previous known preconsolidation pressure ${p_c}_{0}$ as follows:
\begin{equation}
  S=\frac{\langle F \left( p_{0}, {{p_c}_{0}} \right) \rangle}{\mu} \, , %= \frac{1}{\mu}(p_0 -{p_c}_{0}) =  \frac{1}{\mu}(p_0 - p_r(\text{OCR}_0)) ,
\end{equation}
Assuming frozen coeffcients at the current increment after linearization, we state the solution of~\eqref{eq:consistenc_multiax} in exponential form as:
\begin{equation}\label{eq:lambdapunto}
  \dot{\lambda} = \left(\dot{\lambda}_{0} + \frac{a}{b}\right) e^{-\frac{b}{c}t} - \frac{a}{b} \,,
\end{equation}
where the coefficients correspond to the following expressions:
\begin{align}\label{eq:lambdapunto_coef}
    a  =  {\textbf{N}} : \mathbb{C}^{e}: \dot{\varepsilon} \, ,&&
    b  =  - {\textbf{N}} : \mathbb{C}^{e}: {\textbf{N}} -  H N_\text{v}   \, ,&&
    c  =  -S \, ,
\end{align}
and $\dot{\lambda}_0 = \dot{\lambda}(t=0)$. Analyzing~\eqref{eq:lambdapunto}, we can see that in the limit when $t \rightarrow \infty$, we recover the plastic multiplier along the lines of elastoplasticity. We then replace the visco-plastic multiplier in the incremental stress in the following way:
\begin{equation}\label{eq:constitutive}
  \begin{aligned}
    \dot{\sigma}_{ij} &= \mathbb{C}^{e}_{ijkl} \left( \dot{\varepsilon}_{kl} - \dot{\varepsilon}^{vp}_{kl} \right) = \mathbb{C}^{e}_{ijkl} \left( \dot{\varepsilon}_{kl} - \dot{\lambda} {\textbf{N}}_{kl}  \right)  \\
    & = \mathbb{C}^{e}_{ijkl}  \dot{\varepsilon}_{kl} - \mathbb{C}^{e}_{ijkl} \left[ \left(\dot{\lambda}_{0} + \frac{a}{b}\right) e^{-\frac{b}{c}t} - \frac{a}{b} \right] {\textbf{N}}_{kl} \\
    & =  \mathbb{C}^{e}_{ijkl}  \dot{\varepsilon}_{kl} - \mathbb{C}^{e}_{ijkl} {\textbf{N}}_{kl} \dot{\lambda}_{0} e^{-\frac{b}{c}t} -  \frac{\mathbb{C}^{e}_{ijmn} \textbf{N}_{mn} \textbf{N}_{pq} \mathbb{C}^{e}_{pqkl} }{{ \textbf{N}}_{ij} \mathbb{C}^{e}_{ijkl} \textbf{N}_{kl} + H N_\text{v} } \left(1 - e^{-\frac{{ \textbf{N}}_{ij} \mathbb{C}^{e}_{ijkl} \textbf{N}_{kl} + H N_\text{v} }{S} t } \right) \dot{\varepsilon}_{kl} \,.
  \end{aligned}
\end{equation}

Finally, assuming a virgin initial state,  $\dot{\lambda}_{0} = 0$, the tangent visco-plastic constitutive tensor reads:
\begin{equation}\label{eq:sigmadot}
  \dot{\sigma}_{ij} = \mathbb{C}^{vp}_{ijkl} \dot{\varepsilon}_{kl} \,,
\end{equation}
with
\begin{equation}\label{eq:tgteconstitutive_0}
  \mathbb{C}^{vp}_{ijkl}  =   \mathbb{C}^{e}_{ijkl}  -  \mathbb{C}^{d}_{ijkl}, \,\text{with} \,\mathbb{C}^{d}_{ijkl} = \frac{\mathbb{C}^{e}_{ijmn} \textbf{N}_{mn} \textbf{N}_{pq} \mathbb{C}^{e}_{pqkl} }{{ \textbf{N}}_{ij} \mathbb{C}^{e}_{ijkl} \textbf{N}_{kl} + H N_\text{v} } \left(1 - e^{-\frac{{ \textbf{N}}_{ij} \mathbb{C}^{e}_{ijkl} \textbf{N}_{kl} + H N_\text{v} }{S} t } \right) \,,
\end{equation}

In~\eqref{eq:tgteconstitutive_0}, when $t \rightarrow 0$, the viscoplastic constitutive tensor tends to the elastic one 
$$\mathbb{C}^{vp}_{ijkl} \rightarrow \mathbb{C}^{e}_{ijkl},
$$ 
whereas when $t \rightarrow \infty$, it tends to the elastoplastic one $\mathbb{C}^{vp}_{ijkl} \rightarrow \mathbb{C}^{ep}_{ijkl}$. Finally, we compute the flow tensor $\textbf{N}_{ij}$ in terms of the stress invariants for a generalized stress state, assuming the modified Cam-Clay yield surface, as follows: 
\begin{equation} \label{eq:Orientation}
  \begin{aligned} \textbf{N}_{ij} &=
    \begin{bmatrix}
      -\frac{1}{3} + \frac{q^2 - 9 p(\sigma'_{11}-p)}{3 (M p)^2} & -\frac{{\sigma'}_{12}}{M^{2} p} & -\frac{{\sigma'}_{13}}{M^{2} p} \\
      -\frac{{\sigma'}_{12}}{M^{2} p} & -\frac{1}{3} + \frac{q^2 - 9 p(\sigma'_{22}-p)}{3 (M p)^2} & -\frac{{\sigma'}_{12}}{M^{2} p}  \\
      -\frac{{\sigma'}_{13}}{M^{2} p} &  -\frac{{\sigma'}_{23}}{M^{2} p} & \frac{1}{3} + -\frac{q^2 - 9 p(\sigma'_{33}-p)}{3 (M p)^2} \\
    \end{bmatrix}
  \end{aligned} \,.
\end{equation}
\subsection{Acoustic tensor as a bifurcation indicator}

We use the classical bifurcation hetory, based on the spectral properties of the constitutive tensor~\citep{ rice76, rice80}, which determines the admissibility condition for a discontinuity. Maxwell's restriction formulates that a jump in the strain increment must follow~\citep{ thomas61, rice80}:
\begin{equation}
  \lJump d \boldsymbol{\varepsilon} \rJump = d \boldsymbol{\gamma} \otimes^{s}  \boldsymbol{n} \,,
  \label{44}
\end{equation}
where $\boldsymbol{n}$ represents the unit vector normal to the surface where we evaluate the stress state, and $ d \boldsymbol{\gamma}$ is a vector that defines the discontinuity direction in the localization. Besides, $\lJump d \boldsymbol{\varepsilon} \rJump$ represents a jump in the strain increment between two points located on opposite sides of the discontinuity surface:
\begin{equation}
  d \boldsymbol{\varepsilon}^{+} = d \boldsymbol{\varepsilon}^{-} + \lJump d \boldsymbol{\varepsilon} \rJump \,.
  \label{45}
\end{equation}

The mechanical constitutive law in incremental form is:
\begin{equation}
  d \boldsymbol{\sigma} = \mathbb{C} : d \boldsymbol{\varepsilon} \,,
  \label{46}
\end{equation}
where $\mathbb{C}$ is the tangent constitutive tensor, see~\eqref{eq:tgteconstitutive_0}. Equilibrium along the discontinuity surface$\Gamma$, considering continuity )equilibrium) of the projected stresses, reads:
\begin{equation}
  \lJump d \boldsymbol{T} \rJump = \lJump d \boldsymbol{\sigma} \cdot \boldsymbol{n} \rJump = d \boldsymbol{T}^{+} - d \boldsymbol{T}^{-} = \left( \mathbb{C} : \lJump d \boldsymbol{\varepsilon}  \rJump \right) \cdot \boldsymbol{n} = 0 \,.
  \label{47}
\end{equation}
From~\eqref{44} and the symmetry properties of the constitutive tensor $ \mathbb{C}$, we rewrite~\eqref{47} as:
\begin{equation}
  \left( \mathbb{C} \cdot \boldsymbol{n} \right) \cdot d\boldsymbol{\gamma} \cdot \boldsymbol{n} = \mathbb{Q} \left( \boldsymbol{n} \right) \cdot d\boldsymbol{\gamma} = 0 \,,
  \label{48}
\end{equation}
where $\mathbb{Q} \left( \boldsymbol{n} \right)$ is the acoustic tensor:
\begin{equation}
  \mathbb{Q} \left( \boldsymbol{n} \right) = \boldsymbol{n} \cdot \mathbb{C} \cdot  \boldsymbol{n}= 0 \,;
  \label{49}
\end{equation}
conventionally, we solve an eigenvalue problem to determine non-trivial solutions for $d\boldsymbol{\gamma} \neq 0$ in~\eqref{48} :
\begin{equation}
  \det \left( \mathbb{Q} \left( \boldsymbol{n} \right) \right) =  0 \,.
  \label{50}
\end{equation}
We expand the acoustic tensor in index notation in Appendix~\ref{app:A}.

\subsection{Necessary conditions for strain localization}

Heretofore, we interpret~\eqref{50} as a statement of the condition for inhomogeneous localization~\citep{ Olsson1999, besuelle2001}, which is incomplete. Thus, we add necessary conditions on the strain rate from energetic considerations. For this, we recall Hill's instability condition~\cite{ Hill1958} on the second-order work density $d^2W$ under incremental perturbation:
\begin{equation}\label{eq:work}
  d^2W = 0 \implies \dot{\boldsymbol{\sigma}}:\dot{\boldsymbol{\varepsilon}} = 0 \,.
\end{equation}
We rewrite~\eqref{eq:work} in terms of the elastic strain using index notation; thus, the instability condition becomes:
\begin{equation}\label{eq:e_evp}
  \mathbb{C}^{e}_{ijkl} \left( \dot{\varepsilon}_{kl} - \dot{\varepsilon}^{vp}_{kl} \right) \dot{\varepsilon}_{ij} = 0 \,.
\end{equation}
The above identity is valid for any $\dot{\varepsilon}_{ij}$, regardless of its direction. This feature, along with the positive definiteness of $\mathbb{C}^{e}_{ijkl}$, allows us to state that:
\begin{equation}\label{eq:e_tend_vp}
  \dot{\varepsilon}_{kl} \rightarrow \dot{\varepsilon}^{vp}_{kl}
\end{equation} 
in the localization onset. 

Alternatively, we express the stress rate in~\eqref{eq:work} in terms of the total strain rate and the viscoplastic constitutive tensor as in~\eqref{eq:sigmadot}, as follows:
\begin{equation}\label{eq:eCvpe}
  \dot{\boldsymbol{\sigma}}:\dot{\boldsymbol{\varepsilon}} = \dot{\varepsilon}_{ij} \mathbb{C}^{vp}_{ijkl} \dot{\varepsilon}_{kl}  = 0 \,,
\end{equation} 
and, considering~\eqref{eq:eps_vp} and~\eqref{eq:e_tend_vp}, expression~\eqref{eq:eCvpe} reads:
\begin{equation}
  \dot{\varepsilon}_{ij} \mathbb{C}^{vp}_{ijkl} \dot{\varepsilon}^{vp}_{kl}  =\dot{\lambda} \,\dot{\varepsilon}_{ij} \mathbb{C}^{vp}_{ijkl}  \textbf{N}_{kl} = 0 \,,
\end{equation} 
We then construct a tensorial quantity that becomes an indicator of the localization in a more robust way than the acoustic tensor, especially under isotropic conditions, which reads:
\begin{equation}
	\mathbb{L}_{ij} := \mathbb{C}^{vp}_{ijkl} \textbf{N}_{kl} \,.
\end{equation}	 
Unlike typical approaches such as~\citep{ Olsson1999}, the localization onset occurs when an eigenvalue of $\mathbb{L}$ becomes zero, and the localization direction is parallel to the eigenvector associated with the zero eigenvalue. We illustrate the usefulness of this definition in the following section.

\subsection{Stress scenario analysis}

\begin{figure}[ht!]
  \centering
  \includegraphics[width=0.25\textwidth]{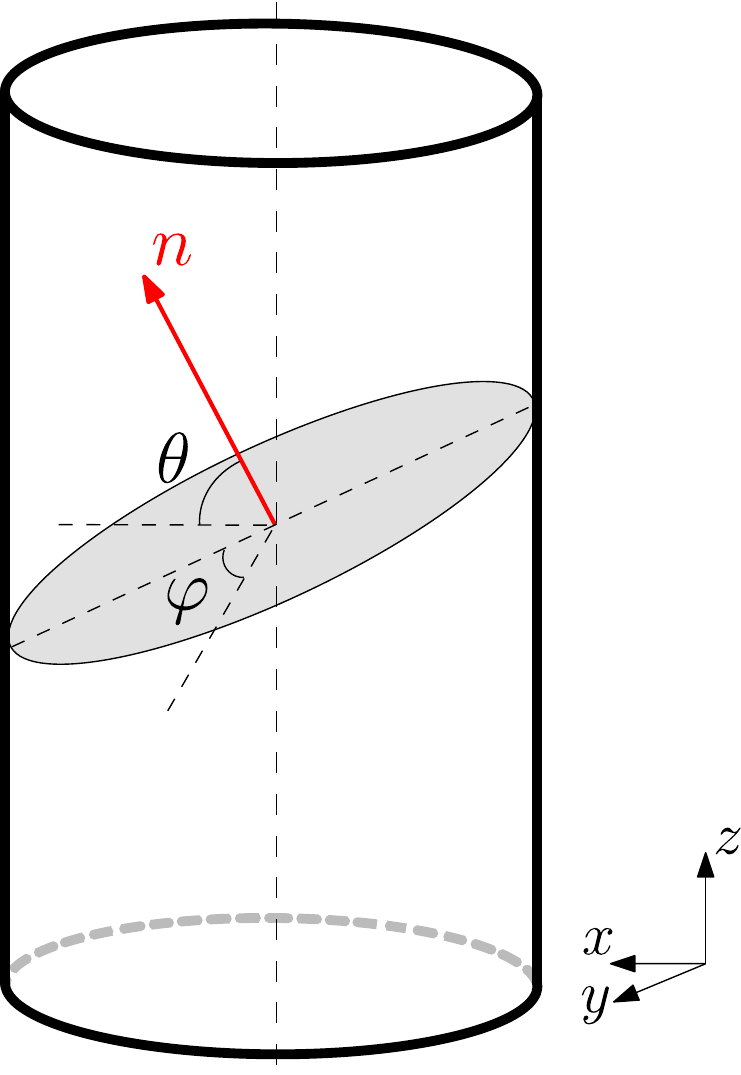}
  \caption{Localization plane with normal direction $\boldsymbol{n}$ expressed by their angular components.}
  \label{fig:normal}
\end{figure}

This section analyzes the localization conditions that trigger the onset of compaction banding for several well-known stress conditions in geomechanics. For this, we express the normal direction  $\boldsymbol{n}$ of the localization plane in terms of two angles $\theta$ and $\varphi$ as follows:
\begin{equation} \label{eq:normal}
  \begin{aligned} \boldsymbol{n} &=
    \begin{bmatrix}
      \cos\theta \sin\varphi \\
      \cos\theta \cos\varphi  \\
      \sin\theta\\
    \end{bmatrix} \,,
  \end{aligned}
\end{equation}
as Figure~\ref{fig:normal} shows. Under triaxial states, we can define an infinite number of localization planes for all values of the angle $\varphi$. We use a fixed value of $\varphi=\pi/2$, which allows us to define the normal $n$ only by the angle $\theta$, measured from the horizontal line.

Using this notation, we can associate the localization planes of the compaction bands that appear in contractive viscoplastic strain regimes with an angle $\theta$ close to $90^\circ$. We analyze the onset of this phenomenon in both compression and extension scenarios under stress-controlled states.

\subsubsection{Isotropic compression/extension}\label{sss:ic}

\proposition{Isotropic pressure states (compression/extension) in isotropic geomaterials localize in all directions $\boldsymbol{n}$ simultaneously.}\label{pp:ic}
\begin{figure}[ht!]
  \centering
  \subfigure[Stress path and plastic flow evolution.]{\includegraphics[width=0.6\textwidth]{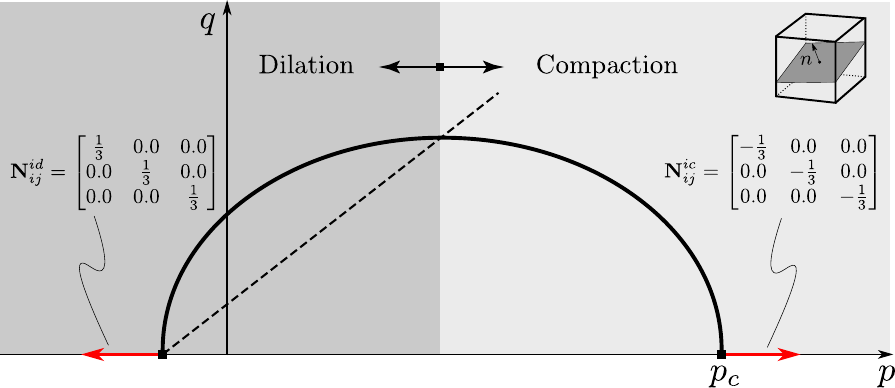} }
  \subfigure[Acoustic tensor degradation.]{\includegraphics[width=0.375\textwidth]{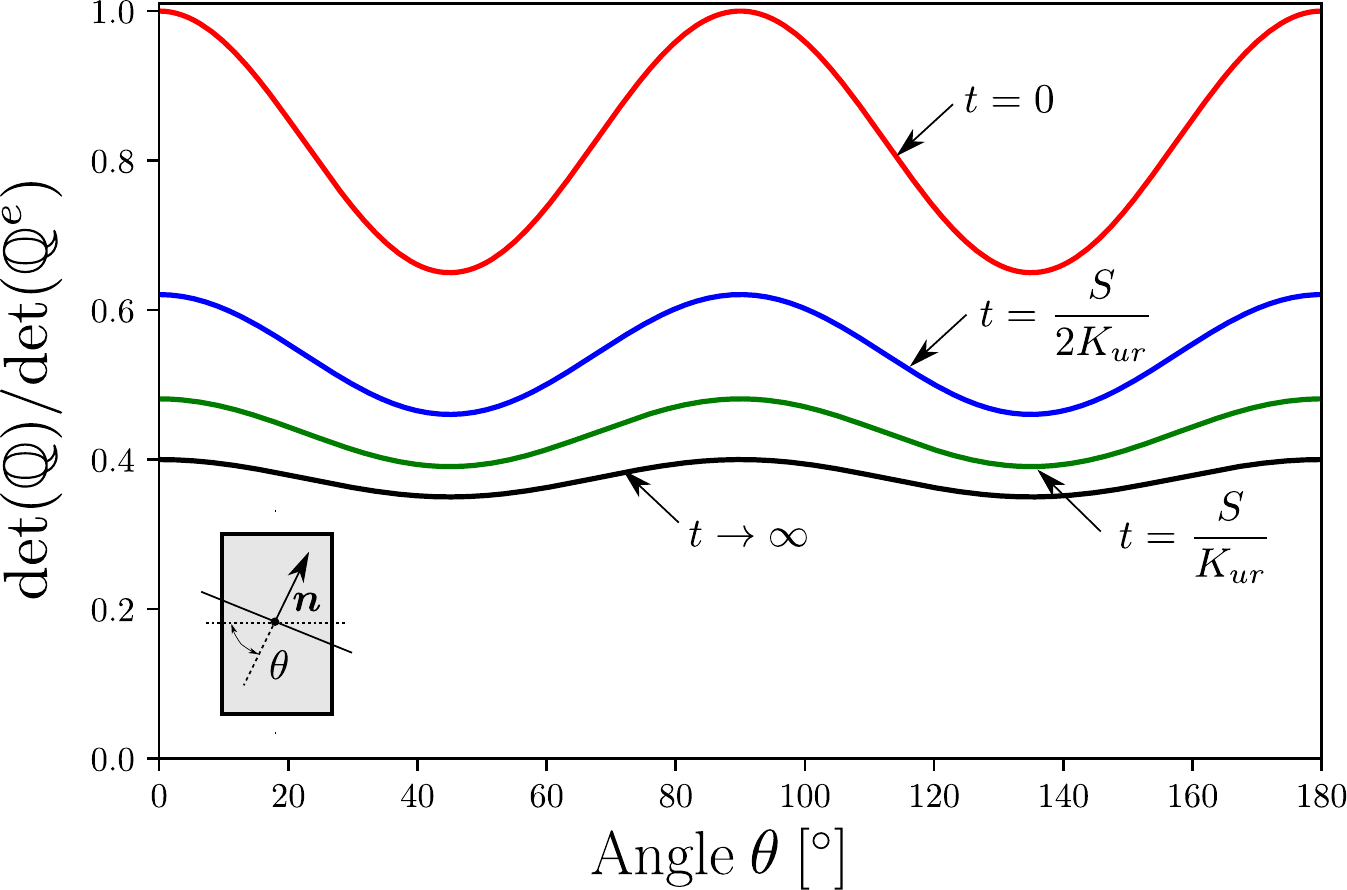} }
  \caption{Isotropic compression/extension.}
  \label{fig:localizationuniaxial}
\end{figure}

\begin{proof}
  The principal stress tensor associated to an isotropic triaxial compression/extension test is:
  \begin{align} \label{eq:sigmaij}
      \begin{bmatrix}
        p' & 0 & 0 \\
        0 & p' & 0 \\
        0 & 0 & p' \\
      \end{bmatrix}
   \qquad\text{with} \qquad{\sigma'}_{11}={\sigma'}_{22}={\sigma'}_{33}=p'\,,
  \end{align}
  Particularizing~\eqref{eq:Orientation} to this stress state, the flow tensor becomes:
  \begin{align} \label{eq:OrientationIC}
      \textbf{N}^{ic}_{ij} =
      \begin{bmatrix}
        -\frac{1}{3} & 0 & 0 \\
        0 & -\frac{1}{3} & 0 \\
        0 & 0 & -\frac{1}{3} \\
      \end{bmatrix}
      &&\text{and} &&
      \textbf{N}^{id}_{ij} =
      \begin{bmatrix}
        \frac{1}{3} & 0 & 0 \\
        0 & \frac{1}{3} & 0 \\
        0 & 0 & \frac{1}{3} \\
      \end{bmatrix} \,,
  \end{align}
  see Figure \ref{fig:localizationuniaxial}(a). Revisiting~\eqref{eq:tgteconstitutive_0}, as the plastic flow direction apprear quadratically in all terms, the resulting instability condition is insensitive to the flow direction its direction (i.e., compression vs extension). Then, assuming a material with bulk modulus $K_{ur}$ and shear modulus $G_{ur}$, the isotropic elastic tensor reads:
  \begin{equation}\label{eq:tgteconstitutiveelastic}
    \mathbb{C}^{e}_{ijkl}  = 
    \begin{bmatrix}
      \frac{4}{3} G_{ur} + K_{ur}  & - \frac{2}{3} G_{ur} +  K_{ur} & - \frac{2}{3} G_{ur} + K_{ur}  & 0 & 0 & 0 \\
      - \frac{2}{3} G_{ur} + K_{ur}  &\frac{4}{3} G_{ur} + K_{ur}  & - \frac{2}{3} G_{ur} + K_{ur} & 0 & 0 & 0\\
      - \frac{2}{3} G_{ur}+  K_{ur}  & - \frac{2}{3} G_{ur} + K_{ur}  &\frac{4}{3} G_{ur} + K_{ur} & 0 & 0 & 0 \\
      0 & 0 & 0 & 2 G_{ur} & 0 & 0 \\
      0 & 0 & 0 & 0 & 2 G_{ur} & 0 \\
      0 & 0 & 0 & 0 & 0 & 2 G_{ur} \\
    \end{bmatrix}  \,,
  \end{equation}
  and computing \eqref{eq:tgteconstitutive_0}, we obtain the following viscoplastic constitutive tensor:
  \begin{equation}\label{eq:tgteconstitutive}
    \mathbb{C}^{vp}_{ijkl}  = G_{ur}
    \begin{bmatrix}
      \frac{4}{3}  + \frac{K_{ur}}{G_{ur}} e^{-\frac{K_{ur}t}{S}} & - \frac{2}{3} + \frac{K_{ur}}{G_{ur}} e^{-\frac{K_{ur}t}{S}}  & - \frac{2}{3}  + \frac{K_{ur}}{G_{ur}} e^{-\frac{K_{ur}t}{S}} & 0 & 0 & 0 \\
      - \frac{2}{3}  + \frac{K_{ur}}{G_{ur}} e^{-\frac{K_{ur}t}{S}} &\frac{4}{3}   + \frac{K_{ur}}{G_{ur}} e^{-\frac{K_{ur}t}{S}} & - \frac{2}{3}  + \frac{K_{ur}}{G_{ur}} e^{-\frac{K_{ur}t}{S}} & 0 & 0 & 0\\
      - \frac{2}{3}  +  \frac{K_{ur}}{G_{ur}} e^{-\frac{K_{ur}t}{S}} & - \frac{2}{3}  + \frac{K_{ur}}{G_{ur}} e^{-\frac{K_{ur}t}{S}} &\frac{4}{3}   + \frac{K_{ur}}{G_{ur}} e^{-\frac{K_{ur}t}{S}} & 0 & 0 & 0 \\
      0 & 0 & 0 & 2  & 0 & 0 \\
      0 & 0 & 0 & 0 & 2 & 0 \\
      0 & 0 & 0 & 0 & 0 & 2  \\
    \end{bmatrix}  \,,
  \end{equation}
  or in terms of Poisson ratio $\frac{K_{ur}}{G_{ur}} = \frac{2 (1+\nu)}{3 (1-2\nu) }$. If $t\rightarrow \infty$, matrix~\eqref{eq:tgteconstitutive} becomes:
  \begin{equation}
    \mathbb{C}^{vp}_{ijkl} \vert_{t\rightarrow \infty}  = G_{ur}
    \begin{bmatrix}
      \frac{4}{3}   & - \frac{2}{3}  & - \frac{2}{3}   & 0 & 0 & 0 \\
      - \frac{2}{3}   &\frac{4}{3}    & - \frac{2}{3}   & 0 & 0 & 0\\
      - \frac{2}{3}  & - \frac{2}{3} &\frac{4}{3}   & 0 & 0 & 0 \\
      0 & 0 & 0 & 2  & 0 & 0 \\
      0 & 0 & 0 & 0 & 2 & 0 \\
      0 & 0 & 0 & 0 & 0 & 2  \\
    \end{bmatrix}  \,.
  \end{equation}
  
  Computing the $\mathbb{L}_{ij}$ tensor for this case, we obtain:
  \begin{equation}
    \mathbb{L}_{ij}  =  
    \begin{bmatrix}
      K_{ur} e^{-\frac{K_{ur}}{S}t}    & 0 & 0     \\
      0   & K_{ur} e^{-\frac{K_{ur}}{S}t}   & 0 \\
      0  & 0 & K_{ur} e^{-\frac{K_{ur}}{S}t}  
    \end{bmatrix}  \,,
  \end{equation}
  where the eigenvalues $\rho_{i}$ and eigenvectors $\boldsymbol{n}_i$ that represent the localization are:
  \begin{equation}
    \rho_{i} = K_{ur} e^{-\frac{K_{ur}}{S}t}, \,\,\text{and} \,\,\boldsymbol{n}_i = K_{ur} e^{-\frac{K_{ur}}{S}t} \,e_{i} \,.
  \end{equation}
  
  The eigenvalues show that the second-order energy density  $d^{2}W$ becomes zero for large time (${t \rightarrow \infty}$). Besides, the localization occurs in all directions $\boldsymbol{n}_i$. Thus, isotropic loadings do not have a preferential localization direction, regardless of the sample drainage (drained or undrained). When a specimen is subject to isotropic pressure, the degradation is only induced on the volumetric stiffness until it completely vanishes, localizing in all directions simultaneously. Then, no localization occurs; the material collapses due to the complete loss of volumetric stiffness. This analysis also reveals that the acoustic tensor in~\eqref{49} is unable to detect this volumetric failure. Figure~\ref{fig:localizationuniaxial}(b) displays this deficiency by showing that the determinant of the acoustic tensor degrades uniformly up to a value larger than zero as time grows (${t\rightarrow \infty}$).
\end{proof}

\subsubsection{Drained triaxial compression}

\proposition{
  In drained triaxial compression tests, the compaction band occurs when the stress path reaches a well-defined point on the yield surface where the plastic flow $\textbf{N}$ is parallel to the maximum principal stress. Mathematically, the plastic flow components meet the condition $\textbf{N}_{22} \rightarrow 0$, $\textbf{N}_{33} \rightarrow 0$ and $\textbf{N}_{11} < 0$.}\label{pp:drained_tx}
\begin{proof}
  Assuming a reference pressure $p_r \leq p_c$ (overconsolidated sample), the stress tensor associated with a deviatoric stress of a drained triaxial compression test reads:
  \begin{equation} \label{eq:sigmaij}
    \begin{aligned} \sigma'_{ij} &=
      \begin{bmatrix}
        3p' - 2p_r & 0 & 0 \\
        0 & p_r & 0 \\
        0 & 0 & p_r \\
      \end{bmatrix}
    \end{aligned} \,,
  \end{equation}
  with $p'$ being the effective isotropic pressure of the sample. Particularizing~\eqref{eq:Orientation} to the stress state, and assuming that $q=3(p'-p_r)$ in triaxial compression, the flow direction in this case is:
  % \begin{equation} \label{eq:OrientationTx}
  % \begin{aligned} \textbf{N}_{ij} &=
  % \begin{bmatrix}
  %			-\frac{1}{3} + \frac{q^2 - 18 p'(p'-p_r)}{3 (M p')^2} & 0 & 0 \\
  %			0 & -\frac{1}{3} + \frac{q^2 + 9 p'(p'-p_r)}{3 (M p')^2} & 0 \\
  %			0 &  0 & -\frac{1}{3} + \frac{q^2 + 9 p'(p'-p_r)}{3 (M p')^2} \\
  % \end{bmatrix}
  % \end{aligned} \,,
  % \end{equation}
  \begin{equation} \label{eq:OrientationTx}
    \begin{aligned} \textbf{N}_{ij} &=
      \begin{bmatrix}
        -\frac{1}{3} - \frac{3(p'^2-p_r^2)}{(M p')^2} & 0 & 0 \\
        0 & -\frac{1}{3} + \frac{3(p'-p_r)(2p' - p_r)}{(M p')^2} & 0 \\
        0 &  0 & -\frac{1}{3} + \frac{3(p'-p_r)(2p' - p_r)}{(M p')^2} \\
      \end{bmatrix}
    \end{aligned} \,.
  \end{equation}
  As in Section~\ref{sss:ic}, we particularize the elastic and visco-plastic constitutive tangent tensors to compute the acoustic tensor to find a localization trigger.  
  
  First, we analyze a normally consolidated drained triaxial test with $p_r = p_c$, where the initial plastic flow $\textbf{N}^{tx}_{ij}\vert_{t_0}$ is the one from Proposition~\ref{pp:ic}, as Figure~\ref{fig:txNC40}(a) shows. The plastic flow direction that triggers the compaction bands is approximately:
  \begin{equation} \label{eq:OrientationTxtrigger}
    \begin{aligned} \textbf{N}^{tx}_{ij}\vert_{t_l} &=
      \begin{bmatrix}
        -0.76 & 0 & 0 \\
        0 &  0 & 0 \\
        0 &  0 &  0 \\
      \end{bmatrix}
    \end{aligned} \,,
  \end{equation}
  that occurs at time $t_l = 30 \frac{S}{H_p}$. Figure~\ref{fig:txNC40}(b) shows the acoustic tensor degradation as it reaches a localized state (${\det(\mathbb{Q})=0}$) for ${\theta = 90^\circ}$ at time $t_l$. Additionally, $\textbf{N}^{tx}_{11} < 0$ and the plastic multiplier $\dot{\lambda} > 0$, implying that the viscoplastic strain is contractive and the localization corresponds to a compaction band.
  \begin{figure}[ht!]
    \centering
    \subfigure[Stress path and plastic flow evolution.]{\includegraphics[width=0.6\textwidth]{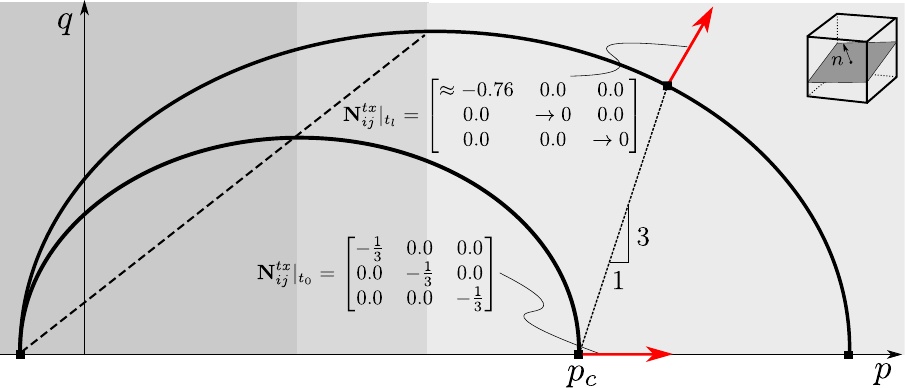} }
    \subfigure[Acoustic tensor degradation.]{\includegraphics[width=0.375\textwidth]{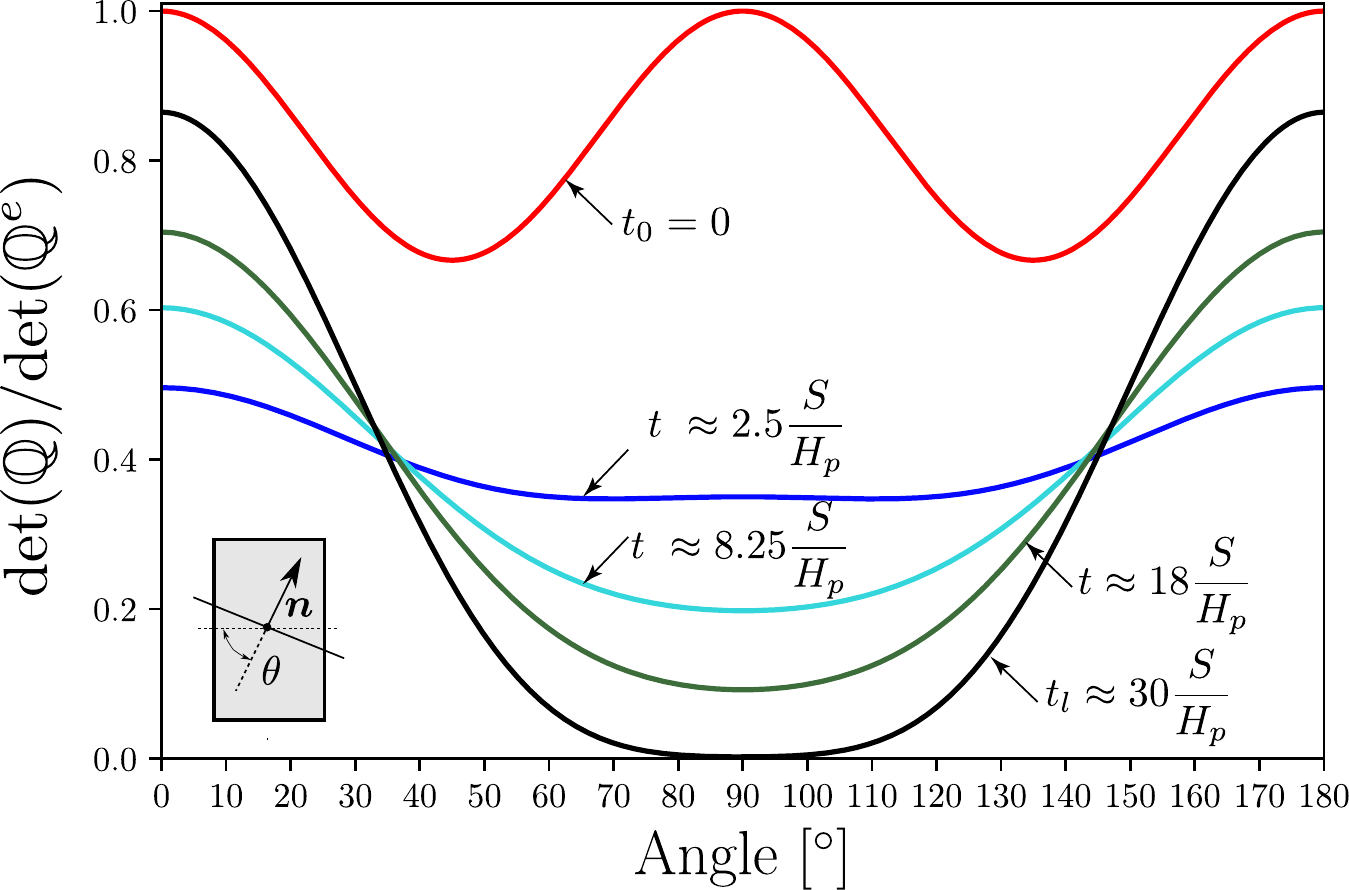} }
    \caption{Drained triaxial compression test: normally consolidated sample ($p_r = p_c$)}
    \label{fig:txNC40}
  \end{figure}
  
  Next, we analyze an overconsolidated sample where ${p_c = 40 \text{MPa}}$ and ${p_r = 22 \text{MPa} < p_c}$. Figure~\ref{fig:txNC22}(a) shows the stress path. The stress path touches the yield surface, inducing the following initial plastic flow:
  \begin{equation} \label{eq:OrientationTxtriggerOverconsol22}
    \begin{aligned} \textbf{N}^{tx}_{ij}\vert_{t_0} &=
      \begin{bmatrix}
        -0.73 & 0 & 0 \\
        0 & -0.03   & 0 \\
        0 &  0 & -0.03  \\
      \end{bmatrix}
    \end{aligned} .
  \end{equation}
  The stress progressively grows and localization happens at time $t_l \approx 0.77 \frac{S}{H_p}$ in the same direction  of~\eqref{eq:OrientationTxtrigger}. In this case, the compaction band appears earlier than in the normally consolidated case.
  
  \begin{figure}[ht!]
    \centering
    \subfigure[Stress path and plastic flow evolution.]{\includegraphics[width=0.55\textwidth]{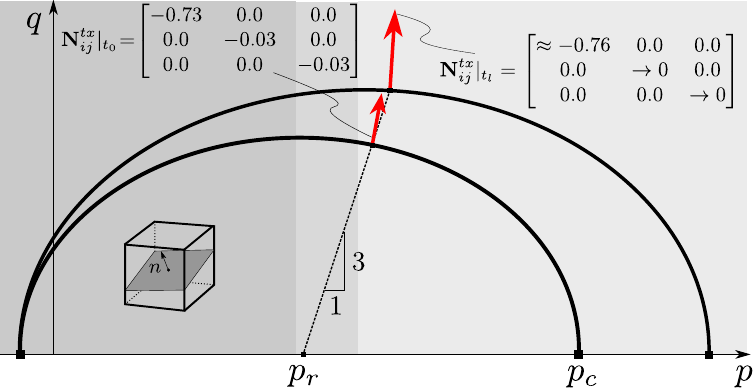} }
    \subfigure[Acoustic tensor degradation.]{\includegraphics[width=0.425\textwidth]{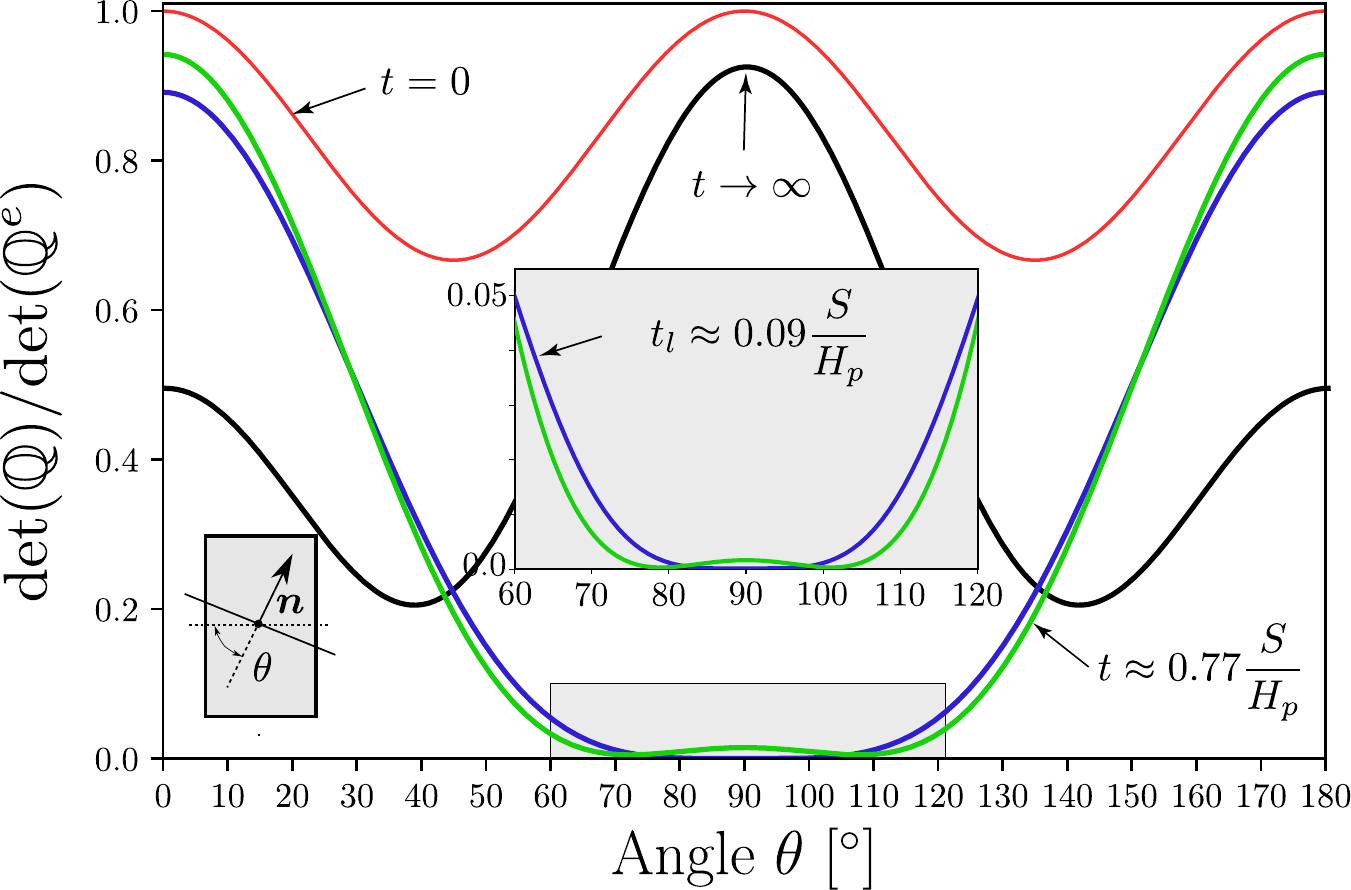} }
    \caption{Drained triaxial compression test: overconsolidated
      sample (reference pressure $p_r = 22 \,\text{MPa}$ \&
      preconsolidation pressure $p_c = 40 \,\text{MPa}$).}
    \label{fig:txNC22}
  \end{figure}
  
  As a consequence, due to the construction of the acoustic tensor and the tangent viscoplastic constitutive tensor, the instability region where the compaction band occurs goes through a point where the plastic flow $\textbf{N}_{ij}$ is parallel to the principal stress applied by the triaxial test. Figure~\ref{fig:Nij} shows the plastic flow evolution in a drained triaxial test for the normally consolidated sample (a) and the over-consolidated one (b). In both cases, the radial components of the plastic flow, ${\textbf{N}_{22}}$ and ${\textbf{N}_{33}}$, become zero when they reach the transition from compressive to extensive regimes, whereas the axial component $\textbf{N}_{11}$ remains negative throughout the whole stress loading history. 
		
  \begin{figure}[ht!]
    \centering
    \subfigure[$\textbf{N}_{ij}$ evolution for the normally consolidated case.]{\includegraphics[width=0.48\textwidth]{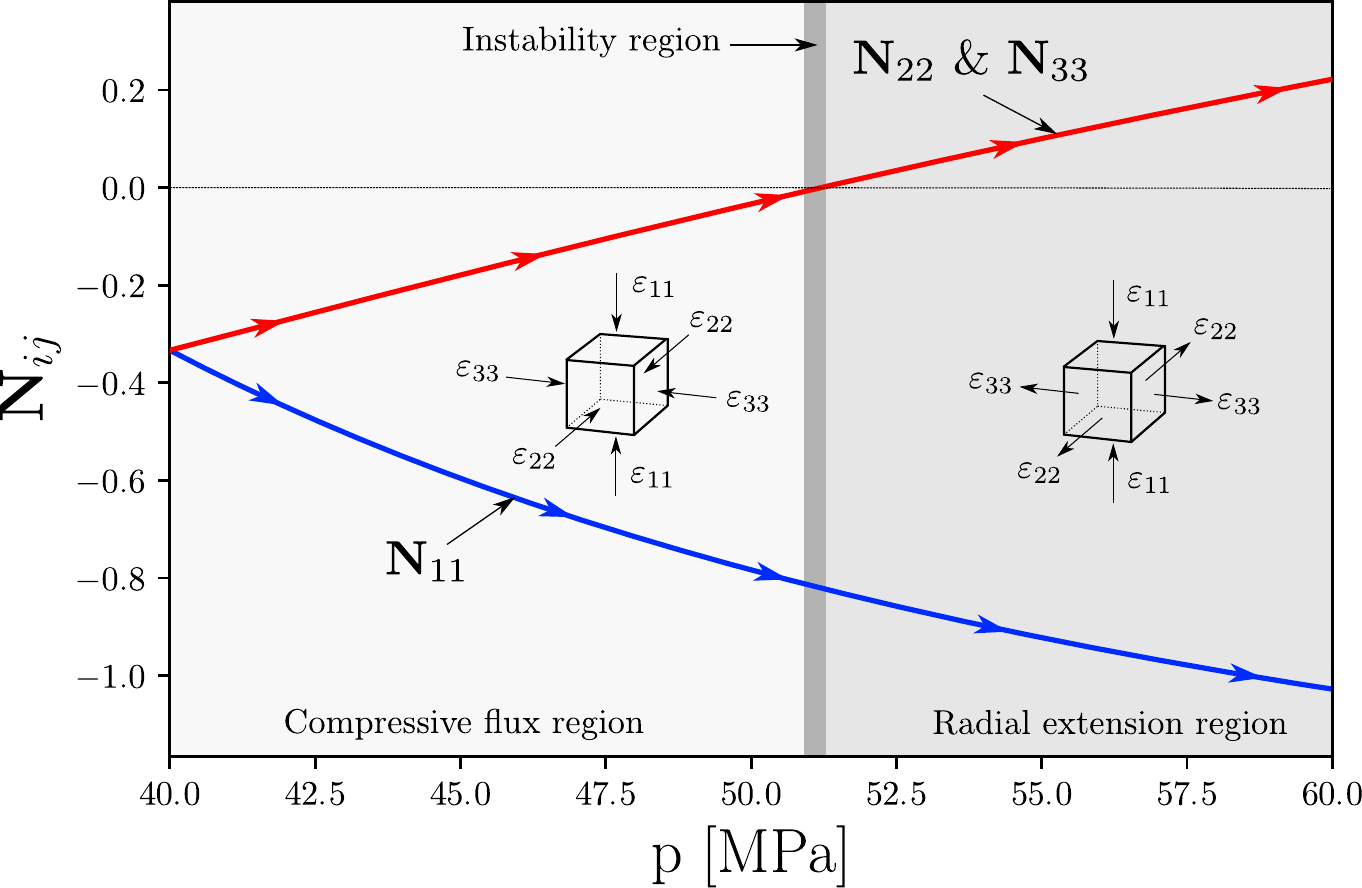} }
    \subfigure[$\textbf{N}_{ij}$ evolution for the over consolidated case.]{\includegraphics[width=0.48\textwidth]{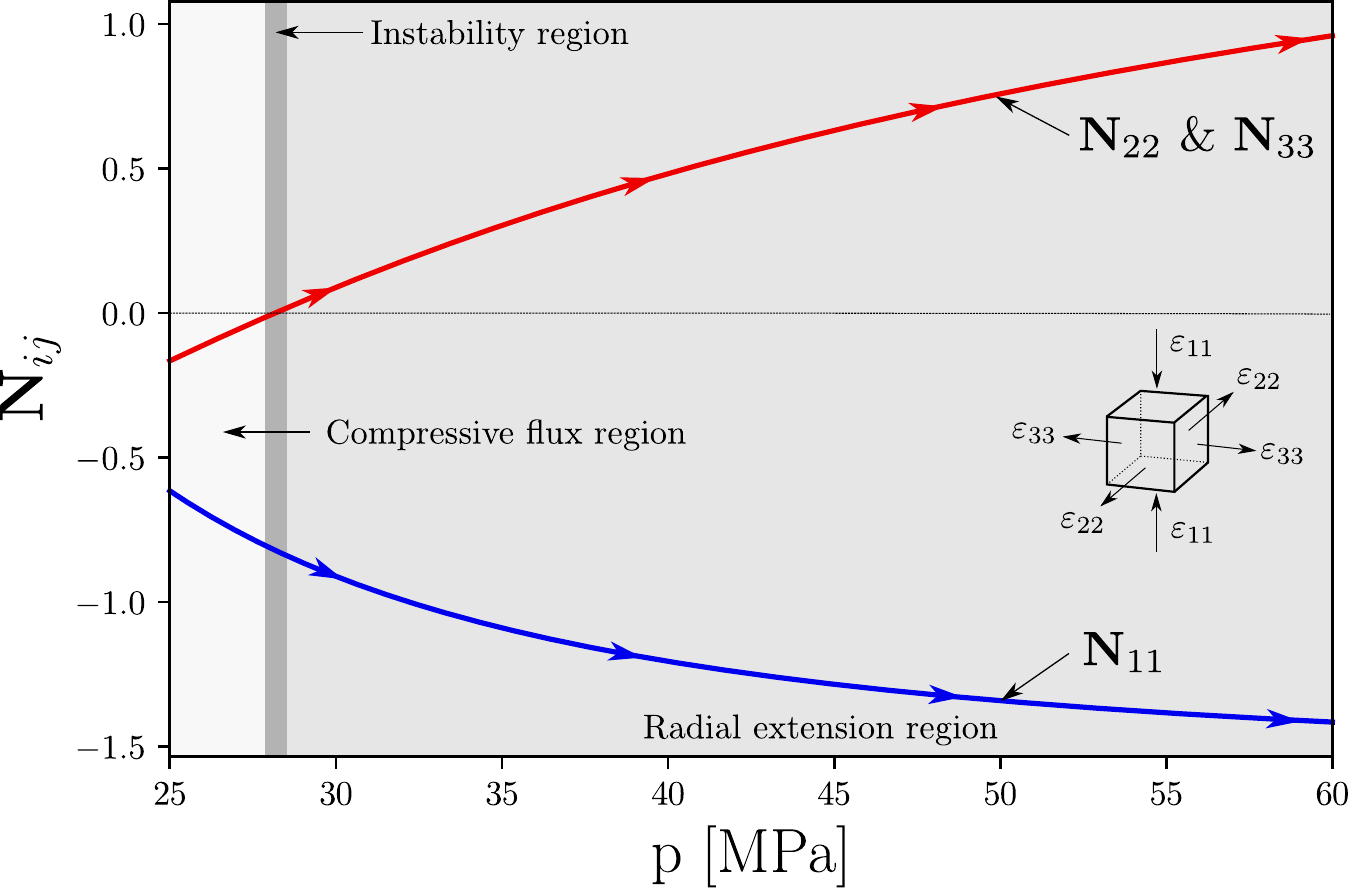} }
    \caption{Plastic flow component evolution and compaction band instability region detection.}
    \label{fig:Nij}
  \end{figure}
\end{proof}

\subsubsection{Drained triaxial extension test}

\proposition{In drained triaxial extension tests, the dilation band is triggered when the stress path touches a well-defined point on the yield surface where the plastic flow $\textbf{N}$ is parallel to the maximum principal stress. Mathematically, plastic flow components meet the conditions $\textbf{N}_{22} \rightarrow 0$, $\textbf{N}_{33} \rightarrow 0$ and $\textbf{N}_{11} > 0$.}

\begin{proof}
  
  Figure~\ref{fig:dilationband}(a) shows the drained triaxial extension test stress path, for an overconsolidated sample, analogous to the previous case. Similarly to the compaction band, the dilation band appears when radial plastic flowes $\textbf{N}_{22} = \textbf{N}_{33} = 0$ under the condition of having a positive plastic flow parallel to the principal stress.  This stress path induces the following initial plastic flow, $\textbf{N}^{dtx}_{ij}\vert_{t_0}$, and localization plastic flow, $\textbf{N}^{dtx}_{ij}\vert_{t_l}$,
  \begin{equation} \label{eq:dilation}
    \begin{aligned} \textbf{N}^{dtx}_{ij}\vert_{t_0} &=
      \begin{bmatrix}
        2.32 & 0 & 0 \\
        0 & -0.55   & 0 \\
        0 &  0 & -0.55  \\
      \end{bmatrix} \quad  \implies \quad 
      \textbf{N}^{dtx}_{ij}\vert_{t_l} &=
      \begin{bmatrix}
        4.57 & 0 & 0 \\
        0 & \rightarrow 0   & 0 \\
        0 &  0 & \rightarrow 0  \\
      \end{bmatrix}
    \end{aligned} \, .
  \end{equation}
  
  Figure~\ref{fig:dilationband}(b) shows that the final condition triggers a localization at $\theta = 90^{\circ}$. However, as $\textbf{N}_{11} > 0$, we observe dilative regime during the localization.  Figure~\ref{fig:dilationband}(c) shows the instability region for this scenario, in terms of the plastic flow in the three directions.

  \begin{figure}[ht!] 
    \centering
    \subfigure[Stress path for triaxial unloading test.]{\includegraphics[width=0.6\textwidth]{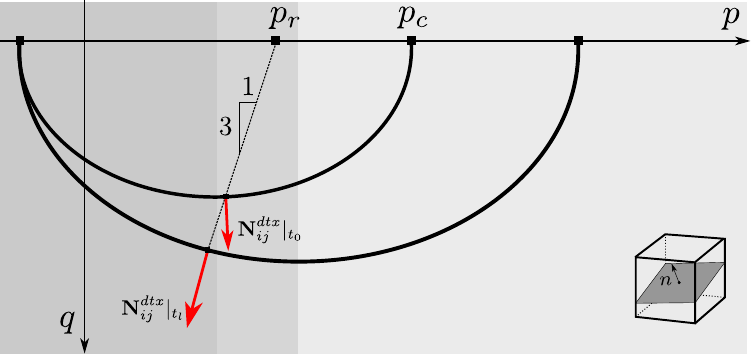} } \\
    \subfigure[Acoustic tensor as a localization inficator]{\includegraphics[width=0.475\textwidth]{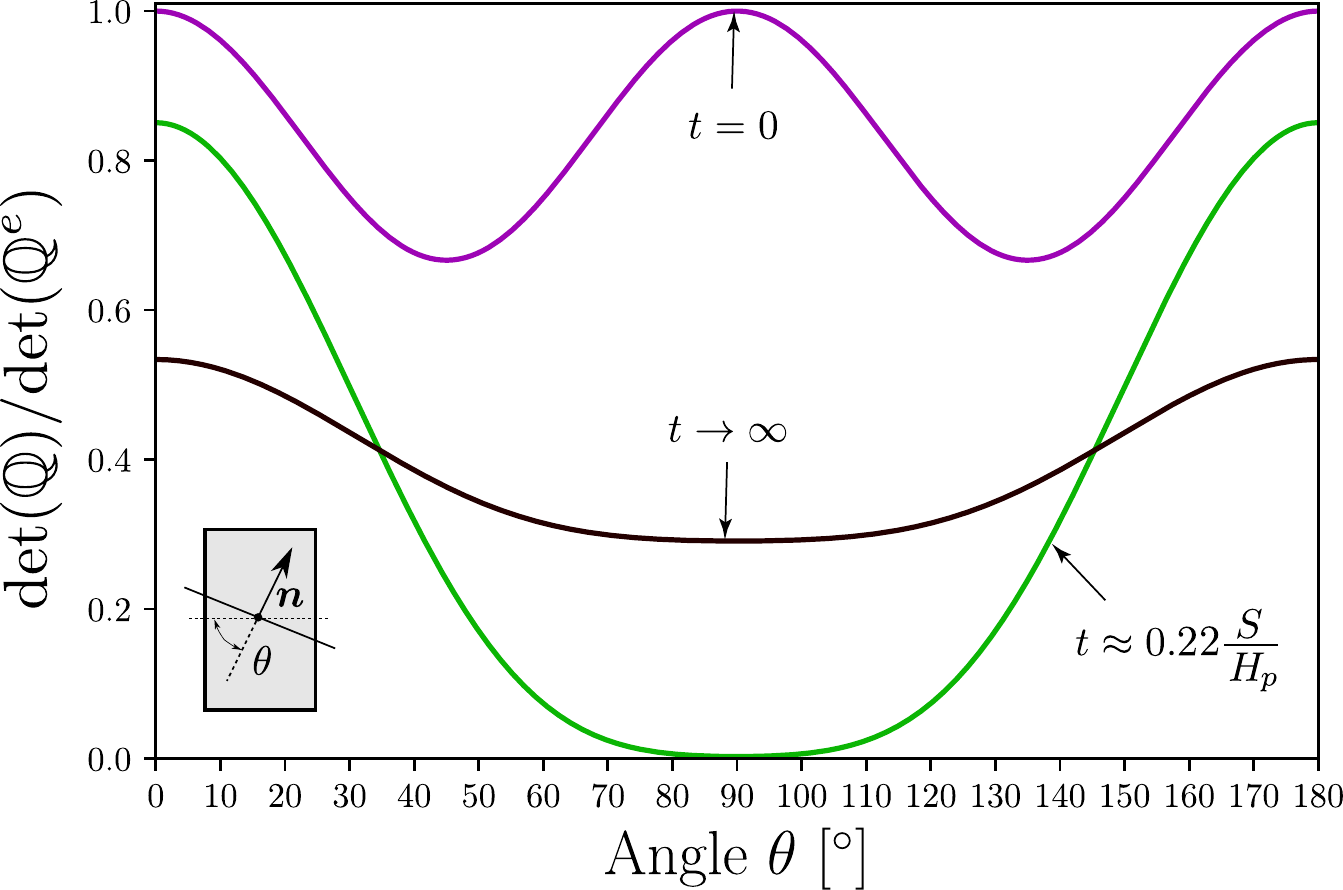} }
    \subfigure[$\textbf{N}_{ij}$ evolution for dilation band formation]{\includegraphics[width=0.475\textwidth]{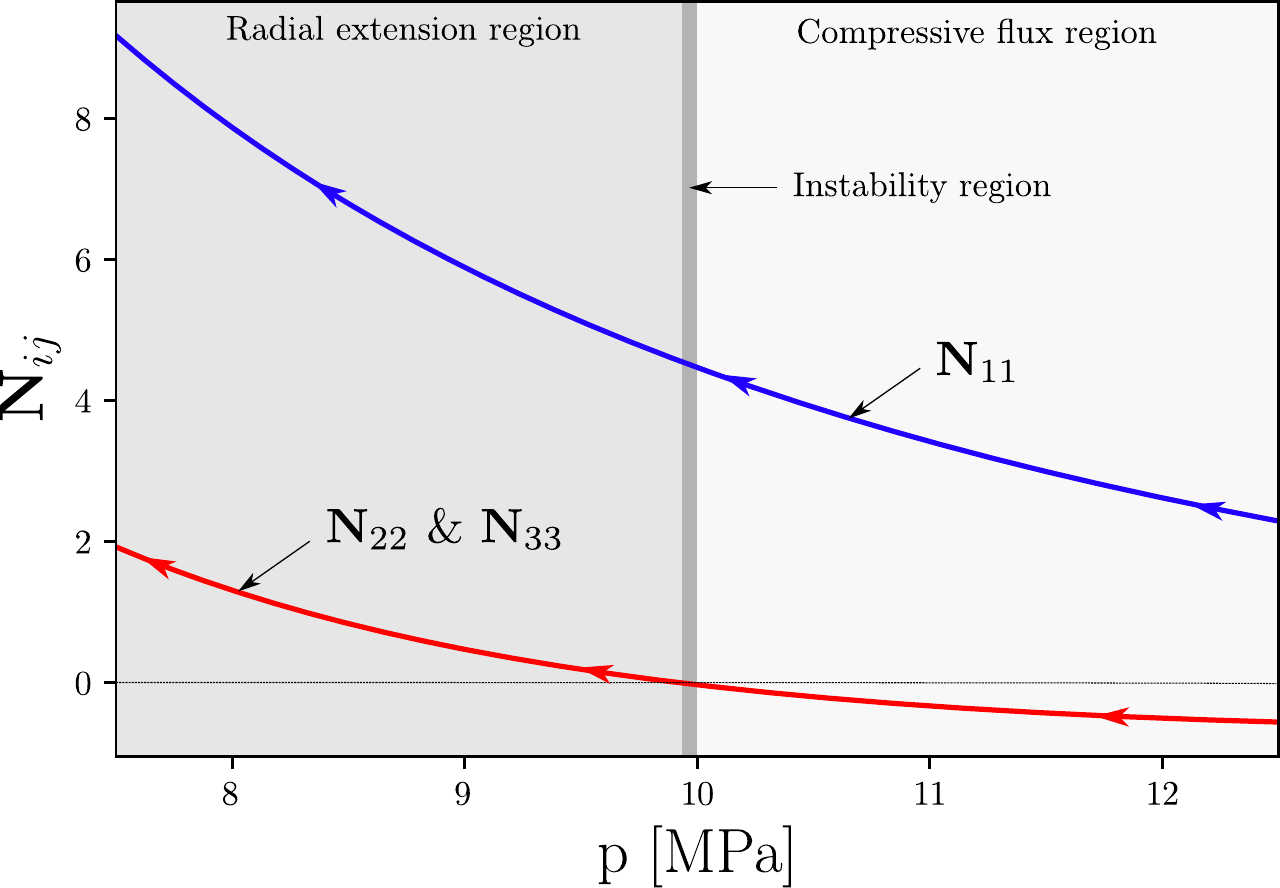} }
    \caption{Dilation band setup}
    \label{fig:dilationband}
  \end{figure}

\end{proof}
 \section{Numerical simulations}\label{section:numexp}

Our analysis framework seeks to explain the appearance of compaction bands in porous rocks processes in several laboratory tests~\citep{ Arroyo2005, Fortin2006, Oka2011, abdallah2021, leuthold2021}. Our numerical experiments induce localization under different triaxial compression conditions, using the Vermeer and Neher's~\citep{ vermeer1999} model, a modified overstress model based on Perzyna's viscoplasticity, that can be understood as a particularization of the constitutive framework of Section~\ref{section:Theory}. Our results show that identical samples subject to different confinement pressures undergo different localization processes. Effectively, the variation of the confinement pressure transitions the localization from shear to compaction bands, as reported in the literature. Our experiments also analyze the bands' periodicity and spacing and their dependence on the material parameters.

\subsection{Constitutive model}

\begin{figure}[ht!]
  \centering
  \includegraphics[width=0.6\textwidth]{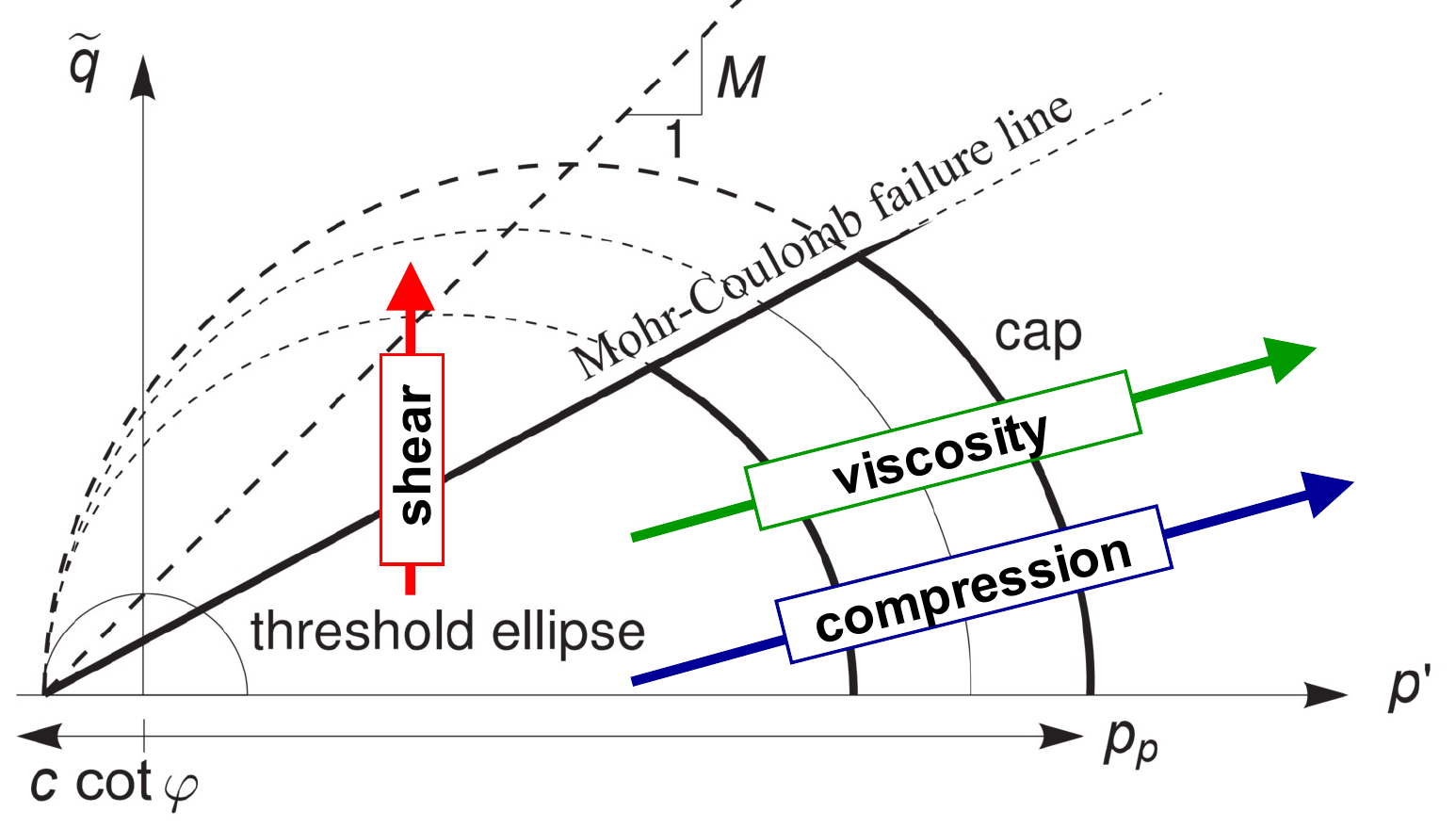} 
  \caption{Vermeer-Neher yield surface and viscous hardening in compression}
  \label{fig:ssc}
\end{figure}

The Vermeer-Neher model~\citep{vermeer1999} incorporates rate-dependent effects into an elastoplastic constitutive model by generalizing the logarithmic creep law for secondary compression~\citep{ bjerrum1967}. In a three-dimensional stress state, the constitutive model combines a perfectly-plastic Mohr-Coulomb yield surface to reproduce shear effects, along with an elliptic cap based on the Modified-Cam Clay (MCC) model~\cite{roscoe1968} that allows simulating the compressive behavior. Moreover, the model incorporates a hardening law that simulates the rate-dependent effect of the material, where all the inelastic strains are considered to be due to creep. Figure~\ref{fig:ssc} shows the yield surface and the viscosity effect associated with the compressive cap. 

%This model uses Roscoe's notation for stresses~\citep{roscoe1958}, that is, for principal effective stresses $\sigma'_1 > \sigma'_2 > \sigma'_3$, the mean and deviatoric stresses can be expressed by:
%\begin{equation}
%	p=\frac{\sigma'_1+\sigma'_2+\sigma'_3}{3} \quad , \quad q=\frac{1}{\sqrt{2}}{\sqrt{(\sigma'_1 - \sigma'_2)^2 + (\sigma'_1 - \sigma'_3)^2 + (\sigma'_2 - \sigma'_3)^2}}.
% \end{equation}

The model introduces the following time-dependent yield function:
\begin{equation}
  F = p^{eq} - p_p^{eq} = p + \frac{q^2}{M^2(p+c \cot\varphi)} - p_p^{eq},
\end{equation}
where $c$ is the cohesion, $\varphi$ is the friction angle, and $M$ is the critical state line's slope (see Figure~\ref{fig:ssc}), ${M = 6\sin\varphi_{cv}/(3-\sin\varphi_{cv})}$, with $\varphi_{cv}$ as the critical-state friction angle. Additionally, the superscript $eq$ represents an equivalent three-dimensional generalization from one-dimensional scenarios where the effective stress ratio $K_0^{NC}$ is known. Thus, for instance, we can compute $p^{eq}$ and $p_p^{eq}$ from the one-dimensional effective stress $\sigma'$ and the preconsolidation pressure $\sigma_p$, respectively.

Assuming standard critical state considerations, this model considers the visco-plastic strain rate evolution entirely in the volumetric part $\dot{\varepsilon}_\text{v}^{vp}$ that extends the one-dimensional creep law, which reads:
\begin{equation}\label{eq:evp_ssc}
  \dot{\varepsilon}_\text{v}^{vp} = -\frac{\mu^*}{\tau}\left(\frac{p^{eq}}{p_p^{eq}}\right)^{\frac{\lambda^*-\kappa^*}{\mu^*}},
\end{equation} 
where $\tau$ is a reference time frame (generally 24 hours), and ${\kappa^*, \lambda^*}$ and $\mu^*$ are indices related to the classical oedometric indices ${C_s, C_c}$ and $C_\alpha$ by:
\begin{align}
  \kappa^* = \frac{2C_s}{2.3(1+e_0)} ,
  && \lambda^* = \frac{C_c}{2.3(1+e_0)} ,
  && \mu^* = \frac{C_\alpha}{2.3(1+e_0)} ,
\end{align}
with $e_0$ the initial void ratio. Finally, we can deduce the preconsolidation pressure $p_p^{eq}$ in~\eqref{eq:evp_ssc} from a MCC state equation modified that accounts for the viscoplastic strain in the following way:
\begin{equation}
	p_p^{eq} = p_{p0}^{eq}\exp\left(-\frac{\varepsilon_\text{v}^{vp}}{\lambda^*-\kappa^*}\right),
\end{equation} 
where $p_{p0}^{eq}$ is an equivalent initial preconsolidation pressure at $t=0$ considering that $\varepsilon_\text{v}^{vp}=0$.

\begin{remark}
  Although the Vermeer and Neher's~\citep{vermeer1999} model was not conceived for modeling rocks, there are examples in the literature where this model is used for this type of geomaterials, especially in subsurface subsidence modeling~\citep{volonte2017, ghisi2021}, given the model's simplicity  and its small number of parameters.
\end{remark}

\subsection{Material parameters selection}

Below, we use standard relationships between different indices to reproduce specific behaviors in this experiment. For our rock, we assume a porosity around 30\%, implying an initial void ratio of $e_0=0.42$. In the literature, compression indices for porous rocks as sandstones typically have values in the range $C_c=0.2-0.4$~\citep{ hupers2012}. In range, we assume $\lambda^*=0.1$. Then, we estimate the other indices based on well-known ratios~\citep{ vermeer1999}. Table~\ref{tab:param_ssc} summarizes the parameters of the model.

\begin{table}[ht!]
  \small
  \centering
  \caption{Model parameters for the modeled rock.}
  \begin{tabular}{l c c c }
    \hline
    \textbf{Parameter}         & \textbf{Symbol} & \textbf{Unit} & \textbf{Value} \\ \hline
    Unit weight &$\gamma$       &   kN/m$^3$    &       22       \\ 
    Compression parameter&$\lambda^*$                &       -       &      0.1       \\ 
    Swelling parameter&$\kappa^*$                 &       -       &      0.01      \\ 
    Creep parameter&$\mu^*$                    &       -       &      5e-4      \\ 
    Poisson's ratio& $\nu_{ur}$ &       -       &      0.15      \\ 
    Cohesion &$c$               &      kPa      &      100       \\ 
    Friction angle &$\varphi$   &   -   &       38$^{\circ}$       \\ 
    Effective stress ratio&$K_0^{NC}$                 &       -       &     0.5239     \\
    Critical state line slope&$M$                        &       -       &     1.563      \\
    Initial preconsolidation pressure&$p'_{p0}$                        &       MPa       &     40       \\ \hline
  \end{tabular}
  \label{tab:param_ssc}
\end{table}

From this parameters selection, we compute a creep ratio (CR), an indirect measure of the viscous contribution in the Vermeer and Neher's~\citep{ vermeer1999} model, as 
$$\text{CR}=\frac{\lambda^*-\kappa^*}{\mu^*}.
$$
Thus, the creep ratio value is $CR=180$, which is small enough to ignore possible rate-dependent effects. However, our numerical examples show that the viscous input induces a change in the strain-localization behavior in our sample.

\subsection{Finite element analysis of triaxial compression tests}

\begin{figure}[ht!]
	\centering
	\includegraphics[width=0.25\linewidth]{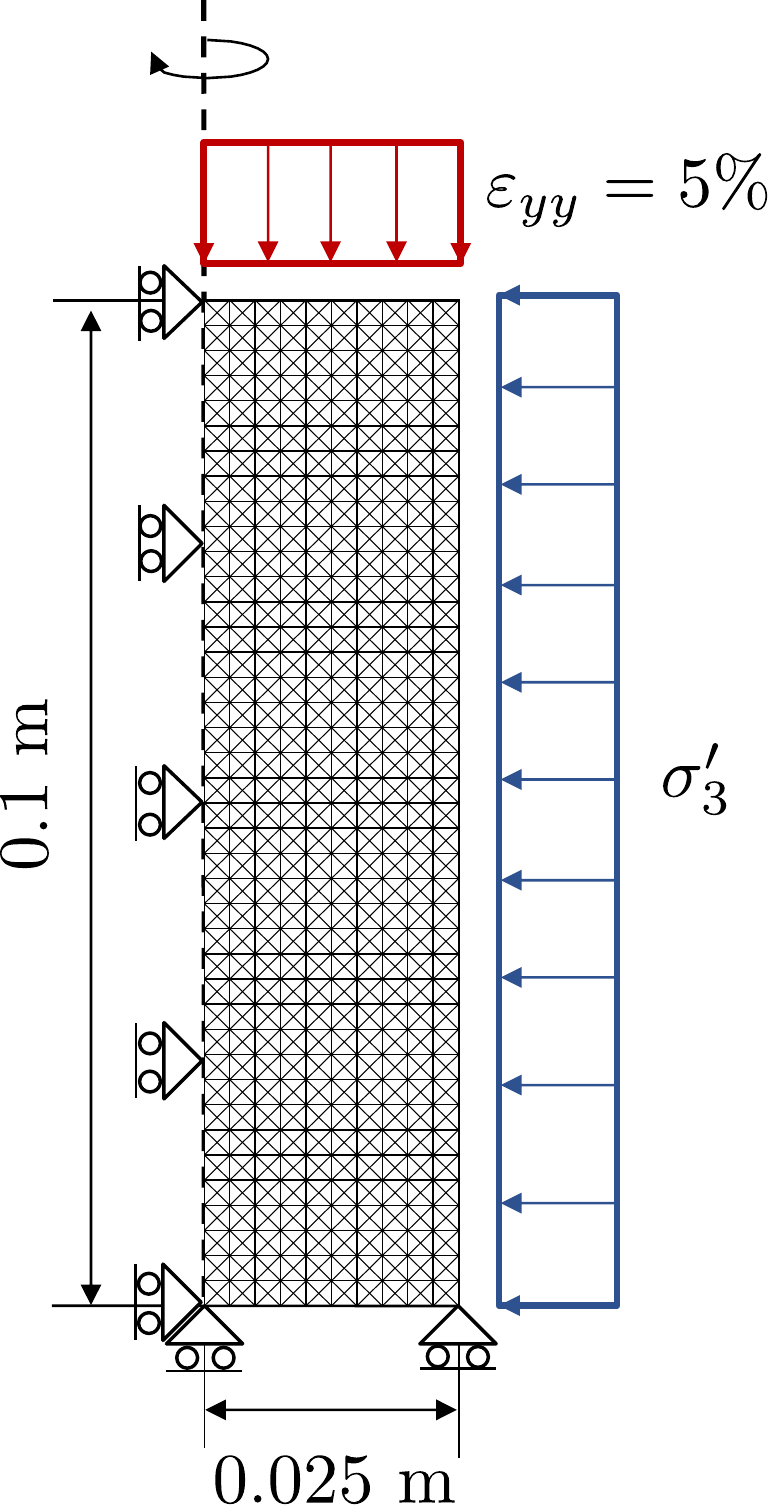}
	\caption{Mesh and boundary conditions for finite element model of shearing stage in a compression triaxial test}
	\label{fig:mesh}
\end{figure}

For the numerical experiments, we employ an axisymmetric strain model for a rectangular domain of ${[0,0.025]~\times~[0,0.1]}~\text{m}^2$, which partitioned into a regular mesh composed of quadratic triangular elements of size $h=0.0025$~m. The boundary conditions are such that displacements normal to $x=0$ and $y =0$ are null. Additionally, we impose a distributed load $\sigma'_3$ at $x=0.1$~m and $y =0.025$~m to simulate the isotropic compression load in the consolidation stage and the confinement pressure during the shearing stage, and a time-dependent displacement $u_y$ at $y =0.1$~m in the shearing stage to reproduce the deviatoric deformation at the top of the sample. In this experiment, the confinement pressure takes values of $\sigma'_3=5,10,14,22$~and~30~MPa, whereas the prescribed displacement is $u_y=5\times 10^{-3}$~m, such that it produces a vertical strain of $\varepsilon_{yy} = 5\%$. Figure~\ref{fig:mesh} sketches the mesh and the boundary conditions considered for the numerical experiments.

We simulate the triaxial compression test using a hydromechanical model that solves the equilibrium and continuity equations similarly to Biot's theory for coupled consolidation. We impose loading strain rate of  ${\dot{\varepsilon}_{yy}= 10^{-5}~\text{s}^{-1}}$, with a time step of $\Delta t = 20$~s. We do not introduce a weak element that induces a localization in the sample for the experiments.

\subsection{Results discussion}

\begin{figure}
  \centering
  \includegraphics[width=0.9\linewidth]{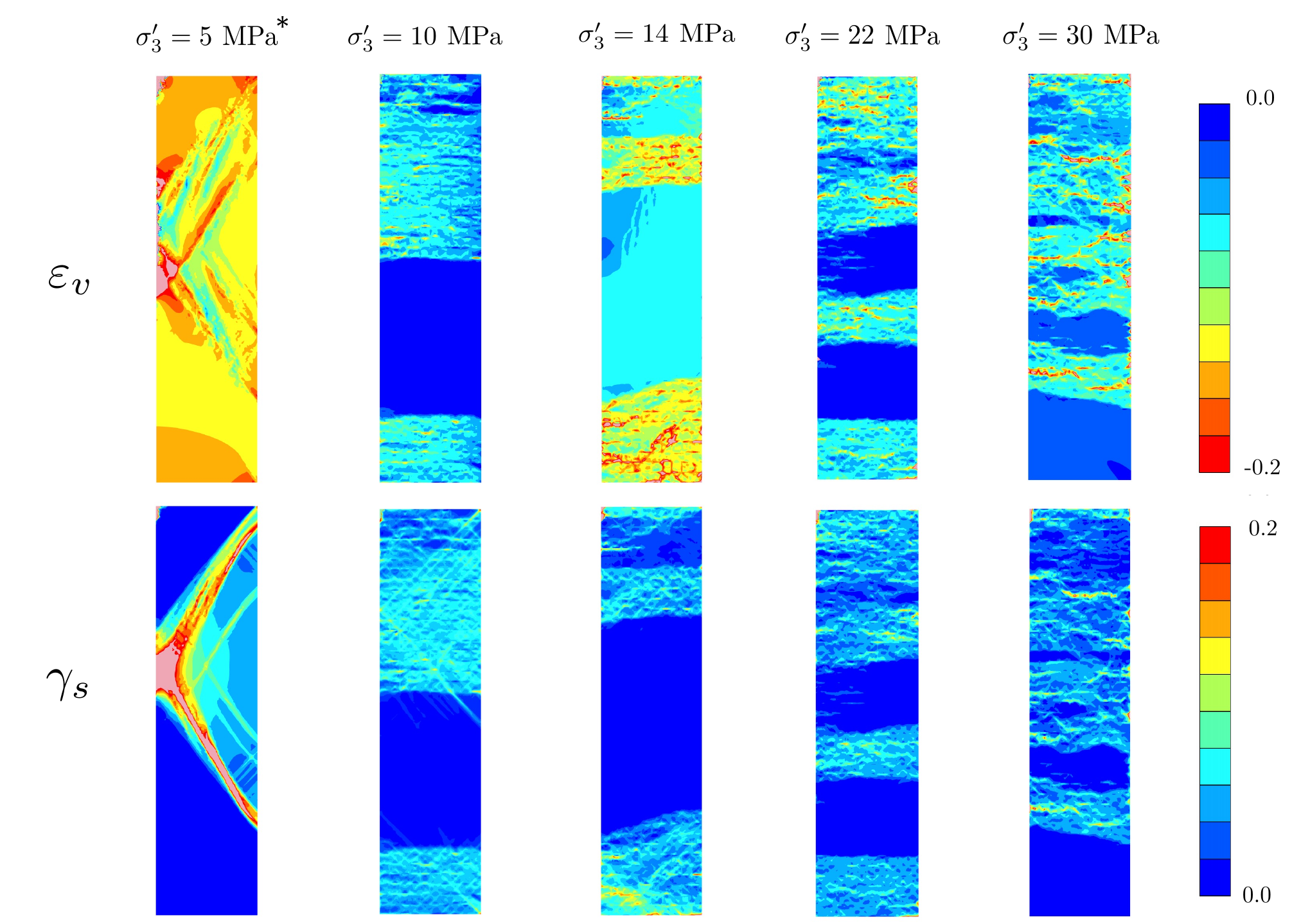}
  \caption{Transition of volumetric ($\varepsilon_\text{v}$) and deviatoric ($\gamma_s$) strain magnitudes as confinement pressure increases. At $\sigma'_3 = 5$~MPa, the values for $\varepsilon_v$ are 20 times smaller than for the rest of cases; thus, the color bar must be read considering this.}
  \label{fig:transition}
\end{figure}

Our results show a transition from shear to compaction banding dependent on the confinement pressure increase, in line with previous works~\citep{ Oka2011}. Figure~\ref{fig:transition} details the volumetric ($\varepsilon_v$) and deviatoric ($\gamma_s$) strain contours for the final condition at each scenario (${\varepsilon_{yy}=5\%}$). The contours for $\sigma'_3=5$~MPa show a well-defined shear band occurring in the specimen, whereas the $\sigma'_3=10$ and 14~MPa cases display a mixed strain localization with some accumulation of both strain components. For the $\sigma'_3=22$ and 30~MPa cases, the compaction banding phenomenon is clear, as well-defined volumetric strain bands in the sample increase in number definition when the confinement pressure increases.

\begin{figure}[ht!]
  \centering
  \includegraphics[width=0.975\linewidth]{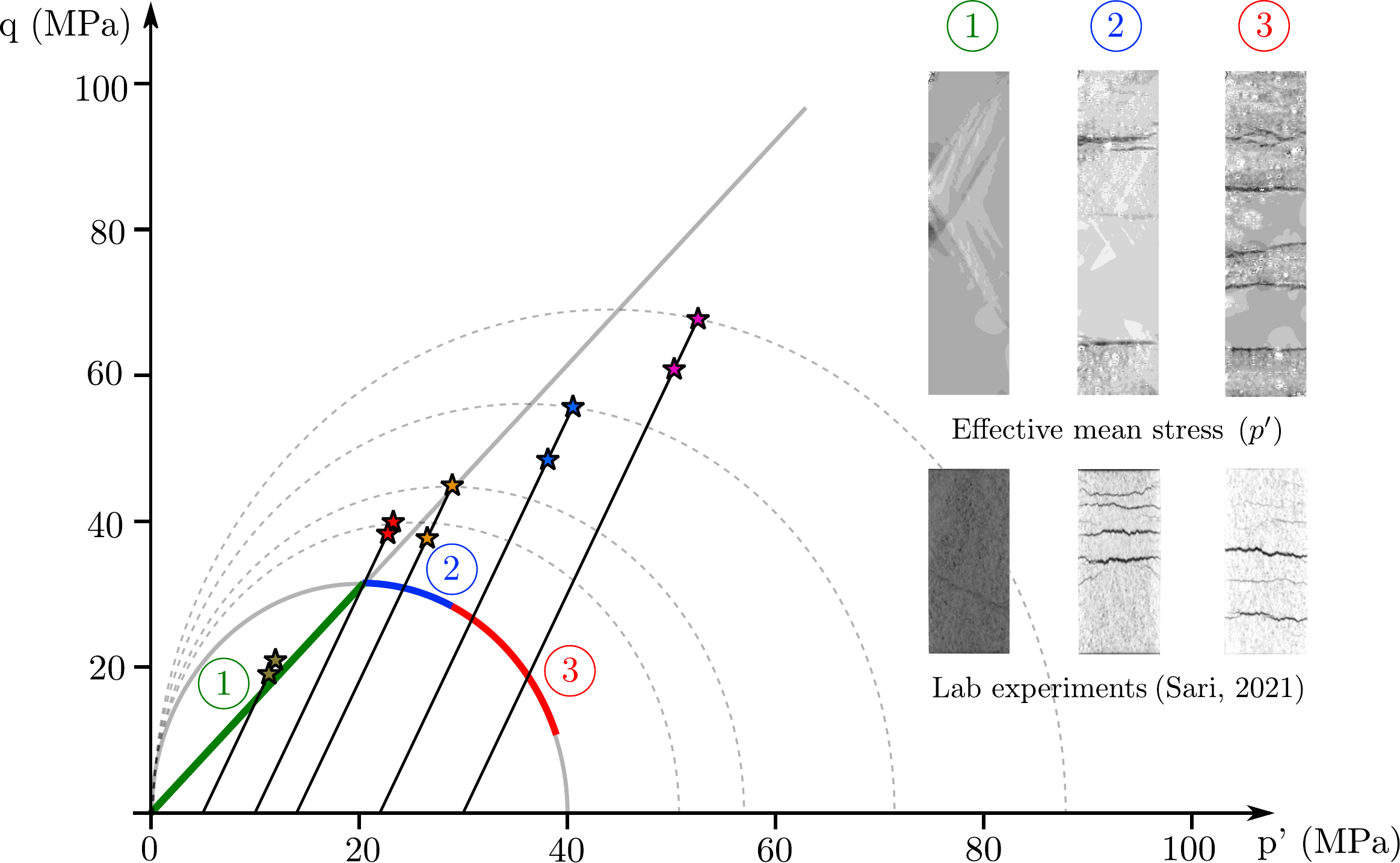}
  \caption{Stress paths for triaxial compression tests for different confinement pressures}
  \label{fig:stress_path}
\end{figure}

We also analyze the transition in the localization behavior through the stress paths from the tests, as Figure~\ref{fig:stress_path} shows. Here, the shear band occurrence (zone \circled{1}) appears for the lowest confinement pressures because the stress path reaches the Mohr-Coulomb yield surface before the cap, implying that the localization is strictly inviscid. For the intermediate confinement pressures, there exists an interplay between the viscous effect produced by pushing the cap (zone \circled{2}), and the failure associated with reaching the Mohr-Coulomb yield surface, producing a compounded (transitional) shear/compaction effect in the sample. Higher confinements generate stress paths that yield a more significant visco-plastic strain inducing the samples to localize purely by compaction (zone \circled{3}). This transition occurs not only through the strain components (see Figure~\ref{fig:transition}) but also through the effective mean stress ($p'$), where the phenomenon evolves from a shear failure, in low confinement pressures, to a well defined and rich set of mean stress accumulation zones in high confinement scenarios. Finally, These results explain experimental observations obtained under similar loading conditions~\citep{ sari2021}, and validate our analytical findings.

%Finally, Figure~\ref{fig:30MPa} shows the two-dimensional model at $\sigma'_3=30$~MPa at the end of the loading stage. Here, we  relate the hydromechanical behavior with the compaction banding phenomenon, as the accumulation zones of preconsolidation pressure and effective stress are associated with the steady-state Darcy velocity magnitude in the sample. While the preconsolidation pressure shows thick accumulation zones, the effective stress exhibits more concentrated localization patterns.

%\begin{figure}[ht!]
%  \centering
%  \includegraphics[width=0.7\linewidth]{30MPa.png}
%  \caption{End of compression test at $\sigma'_3=30$~MPa.}
%  \label{fig:30MPa}
%\end{figure}
 \section{Conclusions}
\label{section:Conclusions}

We present a theoretical and computational description of the strain localization in geomaterials. Specifically, in porous rocks, we describe the process by which a sample with a given preconsolidation pressure under different confinement pressures will either localize in shear or compaction bands. First, we describe the compaction band localization as a bifurcation problem in rate-dependent critical-state materials. The resulting visco-plastic constitutive model is consistent; we construct it from basic assumptions and demonstrate its efficiency in analyzing the bifurcation of homogeneous deformation states in rate-dependent materials. Besides, the spectral analysis of the localization indicator tensor $\mathbb{L}$ overcame the issues associated with the determinant of the classical acoustic tensor $\mathbb{Q}$ under isotropic stress states, allowing us to give a simple description of the localization phenomenon for these cases. We study the onset of compaction bands for well-known stress scenarios in geomechanical tests and establish a series of localization onset conditions, including triggers for both compaction and dilation bands. Then, our numerical experiments validate the compaction banding triggering description from our analytical findings. These simulations particularize our analytical approach for bifurcation; its results confirm how the confinement pressure controls the transition from shear banding to compaction banding. Additionally, these simulation results explain experimental observations carried out under similar loading conditions~\citep{ sari2021}.

%-------------------------------------------------------------------------------%

 \appendix
 \section{Index notation for $\mathbb{Q}$ and $\mathbb{L}$ tensors}
\label{app:A}
Considering index notation, we can express the acoustic tensor in a three-dimensional state as follows:
\begin{align}
	\mathbb{Q}_{jk}  =& \boldsymbol{n}_{i} \ \mathbb{C}_{ijkl} \ \boldsymbol{n}_{l} \\
	\mathbb{Q}_{jk} =& \left( \boldsymbol{n}_{1} \ \mathbb{C}_{1jk1} + \boldsymbol{n}_{2} \ \mathbb{C}_{2jk1} + \boldsymbol{n}_{3} \ \mathbb{C}_{3jk1}  \right) \cdot \boldsymbol{n}_{1} + \left( \boldsymbol{n}_{1} \ \mathbb{C}_{1jk2} + \boldsymbol{n}_{2} \ \mathbb{C}_{2jk2} + \boldsymbol{n}_{3} \ \mathbb{C}_{3jk2} \right) \cdot \boldsymbol{n}_{2}  \\
	& +  \left( \boldsymbol{n}_{1} \ \mathbb{C}_{1jk3} + \boldsymbol{n}_{2} \ \mathbb{C}_{2jk3} + \boldsymbol{n}_{3} \ \mathbb{C}_{3jk3} \right) \cdot \boldsymbol{n}_{3} ,
	\label{51}
\end{align}
where each component is expressed by:
\begin{align}
	&\mathbb{Q}_{11} =  \boldsymbol{n}_{1}  \mathbb{C}_{1111} \boldsymbol{n}_{1} + \boldsymbol{n}_{2} \mathbb{C}_{2112} \boldsymbol{n}_{2}  + \boldsymbol{n}_{3} \mathbb{C}_{3113} \boldsymbol{n}_{3}  =  \boldsymbol{n}_{1}  \mathbb{C}_{11} \boldsymbol{n}_{1} + \boldsymbol{n}_{2} \mathbb{C}_{44} \boldsymbol{n}_{2}  + \boldsymbol{n}_{3} \mathbb{C}_{55} \boldsymbol{n}_{3} \ ,\\
	&\mathbb{Q}_{12} = \boldsymbol{n}_{2}  \mathbb{C}_{2121} \boldsymbol{n}_{1} + \boldsymbol{n}_{1}  \mathbb{C}_{1122} \boldsymbol{n}_{2}  = \boldsymbol{n}_{2}  \mathbb{C}_{44} \boldsymbol{n}_{1} + \boldsymbol{n}_{1}  \mathbb{C}_{12} \boldsymbol{n}_{2}  \ , \\
	&\mathbb{Q}_{13} = \boldsymbol{n}_{3}  \mathbb{C}_{3131} \boldsymbol{n}_{1} + \boldsymbol{n}_{1}  \mathbb{C}_{1133} \boldsymbol{n}_{3} = \boldsymbol{n}_{3}  \mathbb{C}_{55} \boldsymbol{n}_{1} + \boldsymbol{n}_{1}  \mathbb{C}_{13} \boldsymbol{n}_{3},\\
	&\mathbb{Q}_{21} = \boldsymbol{n}_{2}  \mathbb{C}_{2211} \boldsymbol{n}_{1} +  \boldsymbol{n}_{1}  \mathbb{C}_{1212} \boldsymbol{n}_{2} =  \boldsymbol{n}_{2}  \mathbb{C}_{21} \boldsymbol{n}_{1} +  \boldsymbol{n}_{1}  \mathbb{C}_{44} \boldsymbol{n}_{2} ,\\
	&\mathbb{Q}_{22} = \boldsymbol{n}_{1}  \mathbb{C}_{1221} \boldsymbol{n}_{1} + \boldsymbol{n}_{2}  \mathbb{C}_{2222} \boldsymbol{n}_{2} + \boldsymbol{n}_{3}  \mathbb{C}_{3223} \boldsymbol{n}_{3} = \boldsymbol{n}_{1}  \mathbb{C}_{44} \boldsymbol{n}_{1} + \boldsymbol{n}_{2}  \mathbb{C}_{22} \boldsymbol{n}_{2} + \boldsymbol{n}_{3}  \mathbb{C}_{66} \boldsymbol{n}_{3}, \\
	&\mathbb{Q}_{23} =  \boldsymbol{n}_{3}  \mathbb{C}_{3232} \boldsymbol{n}_{2} + \boldsymbol{n}_{2}  \mathbb{C}_{2233} \boldsymbol{n}_{3} =  \boldsymbol{n}_{3}  \mathbb{C}_{66} \boldsymbol{n}_{2} + \boldsymbol{n}_{2}  \mathbb{C}_{23} \boldsymbol{n}_{3}  , \\
	&\mathbb{Q}_{31} =  \boldsymbol{n}_{3}  \mathbb{C}_{3311} \boldsymbol{n}_{1} + \boldsymbol{n}_{1}  \mathbb{C}_{1313} \boldsymbol{n}_{3} =  \boldsymbol{n}_{3}  \mathbb{C}_{31} \boldsymbol{n}_{1} + \boldsymbol{n}_{1}  \mathbb{C}_{55} \boldsymbol{n}_{3} , \\
	&\mathbb{Q}_{32} =  \boldsymbol{n}_{3}  \mathbb{C}_{3322} \boldsymbol{n}_{2} + \boldsymbol{n}_{2}  \mathbb{C}_{2323} \boldsymbol{n}_{3} = \boldsymbol{n}_{3}  \mathbb{C}_{32} \boldsymbol{n}_{2} + \boldsymbol{n}_{2}  \mathbb{C}_{66} \boldsymbol{n}_{3}, \\
	&\mathbb{Q}_{33} =  \boldsymbol{n}_{1}  \mathbb{C}_{1331} \boldsymbol{n}_{1} + \boldsymbol{n}_{2}  \mathbb{C}_{2332} \boldsymbol{n}_{2} + \boldsymbol{n}_{3}  \mathbb{C}_{3333} \boldsymbol{n}_{3} = \boldsymbol{n}_{1}  \mathbb{C}_{55} \boldsymbol{n}_{1} + \boldsymbol{n}_{2}  \mathbb{C}_{66} \boldsymbol{n}_{2} + \boldsymbol{n}_{3}  \mathbb{C}_{33} \boldsymbol{n}_{3} . 
	\label{52}
\end{align}

Then, we derive the components of the $\mathbb{L}_{ij}$ tensor in the following way:
\begin{align}
	\mathbb{L}_{ij}  =&  \mathbb{C}_{ijkl} \  \textbf{N}_{kl} \\
	\mathbb{L}_{ij} =&  \mathbb{C}_{ij11} \  \textbf{N}_{11} + \mathbb{C}_{ij22} \  \textbf{N}_{22} + \mathbb{C}_{ij33} \  \textbf{N}_{33} + \mathbb{C}_{ij12} \  \textbf{N}_{12} + \\
	& \mathbb{C}_{ij13} \  \textbf{N}_{13} + \mathbb{C}_{ij23} \  \textbf{N}_{23} + \mathbb{C}_{ij21} \  \textbf{N}_{21} + \mathbb{C}_{ij31} \  \textbf{N}_{31} + \mathbb{C}_{ij32} \  \textbf{N}_{32},
	\label{51}
\end{align}
thus, the components read:
\begin{align}
	&\mathbb{L}_{11}  =   \mathbb{C}_{11} \textbf{N}_{11} + \mathbb{C}_{12} \textbf{N}_{22} +  \mathbb{C}_{13} \textbf{N}_{33} \ ,\\
	&\mathbb{L}_{12} =  2   \mathbb{C}_{44} \textbf{N}_{12} \ , \\
	&\mathbb{L}_{13} = 2  \mathbb{C}_{55} \textbf{N}_{13},\\
	&\mathbb{L}_{21} = 2   \mathbb{C}_{44} \textbf{N}_{12} ,\\
	&\mathbb{L}_{22} =  \mathbb{C}_{12}  \textbf{N}_{11} + \mathbb{C}_{22} \textbf{N}_{22} +  \mathbb{C}_{13} \textbf{N}_{33}, \\
	&\mathbb{L}_{23} =  2 \mathbb{C}_{66} \textbf{N}_{23}  , \\
	&\mathbb{L}_{31} =  2  \mathbb{C}_{55} \textbf{N}_{13} , \\
	&\mathbb{L}_{32}  = 2  \mathbb{C}_{66} \textbf{N}_{23}  , \\
	&\mathbb{L}_{33} = \mathbb{C}_{13}  \textbf{N}_{11} + \mathbb{C}_{23} \textbf{N}_{22} +  \mathbb{C}_{33} \textbf{N}_{33}. 
	\label{52}
\end{align}

%-------------------------------------------------------------------------------%

%\footnotesize

\bibliographystyle{unsrt}
\bibliography{ref,mybibfile}

\end{document}